\documentclass[aps,twocolumn,prd,superscriptaddress,preprintnumbers,showpacs]{revtex4-1}
\usepackage{graphicx}
\usepackage{bm}
\usepackage{amsmath}
\usepackage{amssymb}
\usepackage{amsthm}

\newcommand{\be}{\begin{equation}}
\newcommand{\ee}{\end{equation}}
\newcommand{\bea}{\begin{eqnarray}}
\newcommand{\eea}{\end{eqnarray}}
\newcommand{\bdm}{\begin{displaymath}}
\newcommand{\edm}{\end{displaymath}}
\newcommand{\beas}{\begin{eqnarray*}}
\newcommand{\eeas}{\end{eqnarray*}}

\begin{document}
\title{Stability of fundamental couplings: a global analysis}

\author{C. J. A. P. Martins}
\email[]{Carlos.Martins@astro.up.pt}
\affiliation{Centro de Astrof\'{\i}sica da Universidade do Porto, Rua das Estrelas, 4150-762 Porto, Portugal}
\affiliation{Instituto de Astrof\'{\i}sica e Ci\^encias do Espa\c co, CAUP, Rua das Estrelas, 4150-762 Porto, Portugal}
\author{A. M. M. Pinho}
\email[]{Ana.Pinho@astro.up.pt}
\affiliation{Centro de Astrof\'{\i}sica da Universidade do Porto, Rua das Estrelas, 4150-762 Porto, Portugal}
\affiliation{Faculdade de Ci\^encias, Universidade do Porto, Rua do Campo Alegre, 4150-007 Porto, Portugal}

\date{27 October 2016}

\begin{abstract}
Astrophysical tests of the stability of fundamental couplings are becoming an increasingly important probe of new physics. Motivated by the recent availability of new and stronger constraints we update previous works testing the consistency of measurements of the fine-structure constant $\alpha$ and the proton-to-electron mass ratio $\mu=m_p/m_e$ (mostly obtained in the optical/ultraviolet) with combined measurements of $\alpha$, $\mu$ and the proton gyromagnetic ratio $g_p$ (mostly in the radio band). We carry out a global analysis of all available data, including the 293 archival measurements of {\it Webb et al.} and 66 more recent dedicated measurements, and constraining both time and spatial variations. While nominally the full datasets show a slight statistical preference for variations of $\alpha$ and $\mu$ (at up to two standard deviations), we also find several inconsistencies between different sub-sets, likely due to hidden systematics and implying that these statistical preferences need to be taken with caution. The statistical evidence for a spatial dipole in the values of $\alpha$ is found at the 2.3 sigma level. Forthcoming studies with facilities such as ALMA and ESPRESSO should clarify these issues.
\end{abstract}

\pacs{04.50.Kd, 98.80.-k}
\maketitle

%%%%%%%%%%%%%%%%%%%%%%%%%%%%%%%%%%%%%%%%%%%%%%%%%%%%%%%%%%%%%%%%%%%%%%%%%%%%%%
\section{\label{intro}Introduction} 

The observational evidence for the acceleration of the universe shows that canonical theories of cosmology and fundamental physics are at least incomplete, and that some currently unknown physics is waiting to be discovered. Tests of the stability of nature's fundamental couplings are becoming an increasingly important component of this search \cite{Uzan,grg}. Very tight constraints exist from local laboratory tests using atomic clocks \cite{Rosenband}, while astrophysical measurements allow a large lever arm which can probe the dynamics of the new degrees of freedom responsible for such putative variations even beyond the regime where these degrees of freedom dominate the cosmological dynamics (that is, deep in the matter era). Furthermore, these measurements---whether they are detections of variations or null results--- constrain Weak Equivalence Principle violations and shed light on the dark energy enigma \cite{WEP,Reconst}. There have been recent indications of possible variations \cite{Dipole}, which are being actively tested.

Direct astrophysical measurements of the fine-structure constant $\alpha$ and the proton-to-electron mass ratio $\mu=m_p/m_e$ are typically carried out in the optical/ultraviolet (there are a few exceptions to this for the $\mu$ case, though only at low redshifts), and up to redshifts which now exceed $z=4$. On the other hand, in the radio/mm band, and typically at lower redshifts, one can measure various combinations of $\alpha$, $\mu$ and the proton gyromagnetic ratio $g_p$. In a recent work \cite{Frigola,Ferreira2015} we carried out a joint statistical analysis of 48 recent dedicated measurements, and highlighted some apparent inconsistencies which could be an indication that systematics may be affecting some of the data.

Since that previous work significant developments occurred. Some of these dedicated measurements have been improved and others have been added, increasing the total to 66, of which 21 are measurements of $\alpha$, 16 of $\mu$ and 29 of several combinations of $\alpha$, $\mu$ and $g_p$. This motivates us to carry out an updated and more thorough consistency analysis, for example dividing the data into several redshift bins. From a theoretical point of view, one may expect different behaviors deep in the matter era and in the more recent dark energy dominated era, with the former allowing larger variations. On the other hand, as already mentioned in the previous paragraphs, different observational techniques are typically used for low and high redshift spectroscopic measurements and these may therefore be vulnerable to different systematics.

Unlike previous works we will also include the 293 archival measurements of Webb {\it et al.} \cite{Dipole}. Although there are some concerns about the quality of the archival data that led to these measurements \cite{Syst}, here we will take them at face value since in any case this dataset provides a useful benchmark with which one can compare the more recent one. Another improvement of our analysis is that we will also use the full set of available measurements to constrain possible spatial variations, updating the analyses in \cite{Dipole,Pinho0}.

Since we are considering variations of several different dimensionless parameters, we should note that in many well-motivated models all such variations will be related in model-dependent ways. For example, in a generic class of unification scenarios discussed in \cite{Campbell,Coc,Luo} the relative variations of these parameters are related via two parameters, one of which is related to the strong sector of the theory while the other is related to the electroweak one. Constraints on these parameters have been obtained in previous works  \cite{Frigola,Ferreira2015,Clocks}, though one finds that current data only constraints a particular combination of these two parameters. Here we will take a simpler (and arguably more intuitive) phenomenological approach, using the data to directly constraining the relation between the relative variation of $\alpha$ and those of the other parameters.

We start in Sect. II by providing an up-to-date list of available dedicated measurements. Sect. III compares the dataset of combined measurements with those of direct measurements of $\alpha$ and $\mu$, while Sect. IV discusses possible time (redshift) variations, Sect. V looks at the relations between the variations of the parameters and Sect. VI focuses on possible spatial variations of $\alpha$. Finally a brief outlook can be found in Sect. VII.

%%%%%%%%%%%%%%%%%%%%%%%%%%%%%%%%%%%%%%%%%%%%%%%%%%%%%%%%%%%%%%%%%%%%%%%%%%%%%%
\section{\label{qsodata}Current spectroscopic measurements}

\begin{figure*}
\begin{center}
\hskip-0.1in
\includegraphics[width=2.6in]{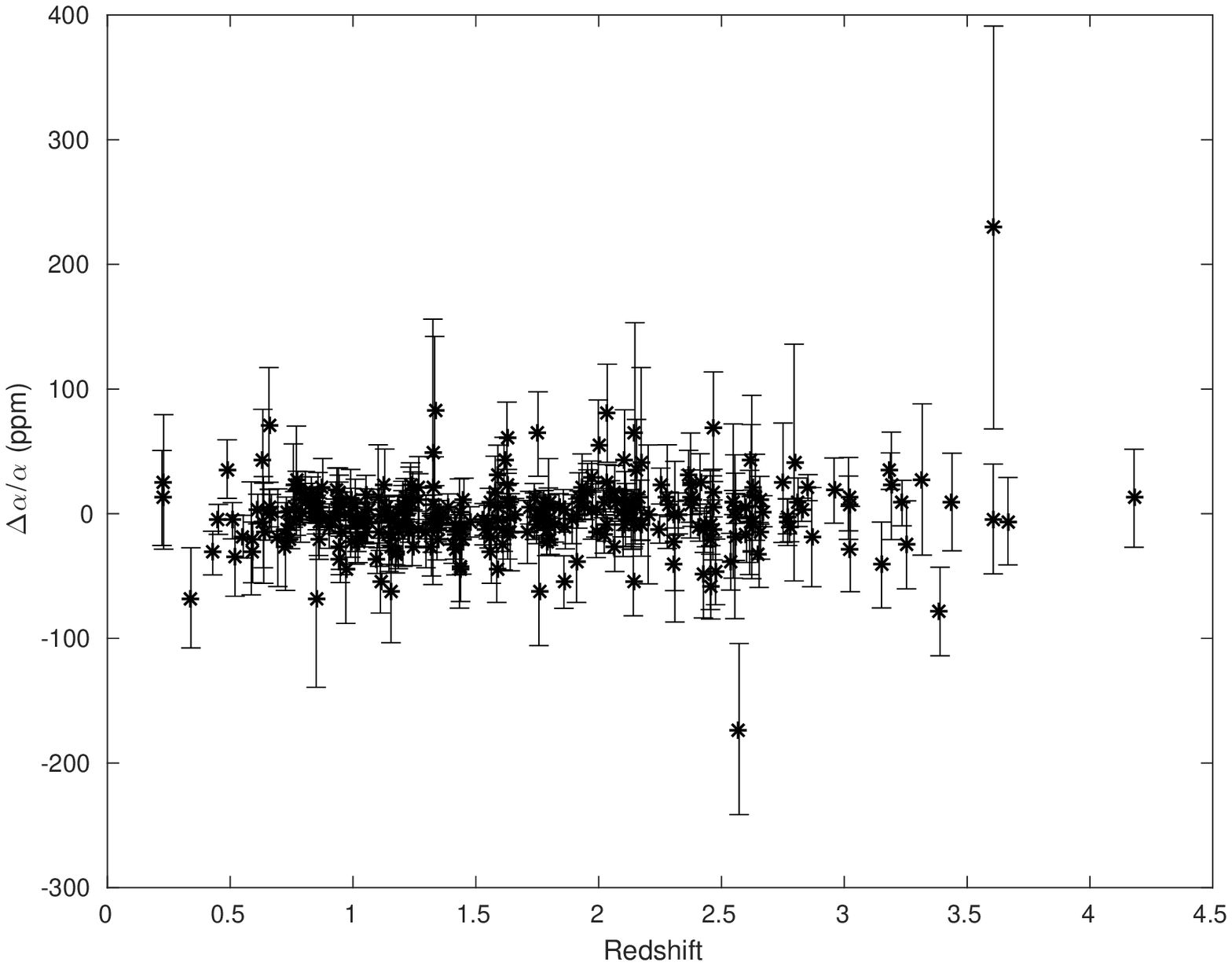}
\hskip0.2in
\includegraphics[width=2.6in]{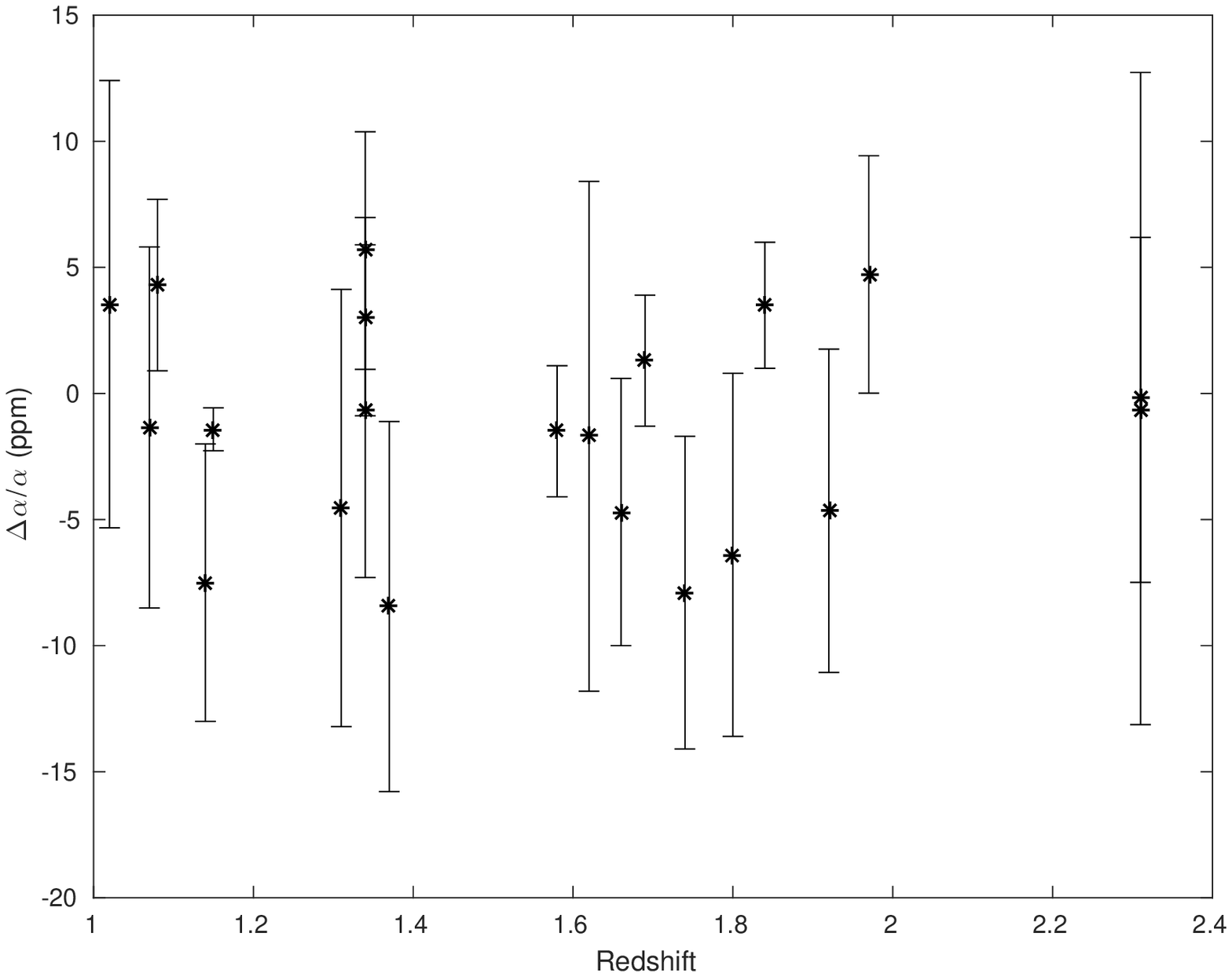}
\vskip0.25in
\hskip-0.1in
\includegraphics[width=2.6in]{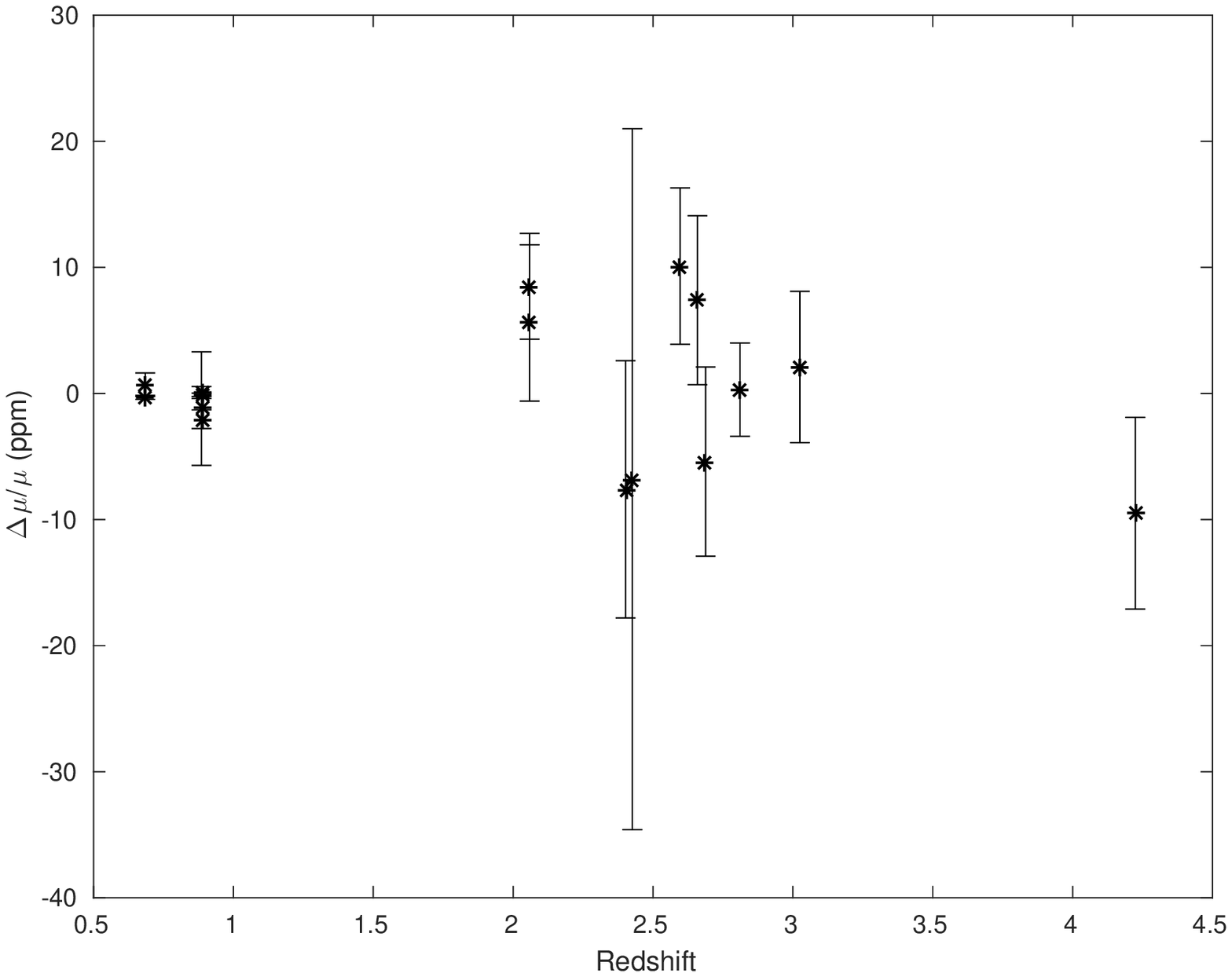}
\hskip0.2in
\includegraphics[width=2.6in]{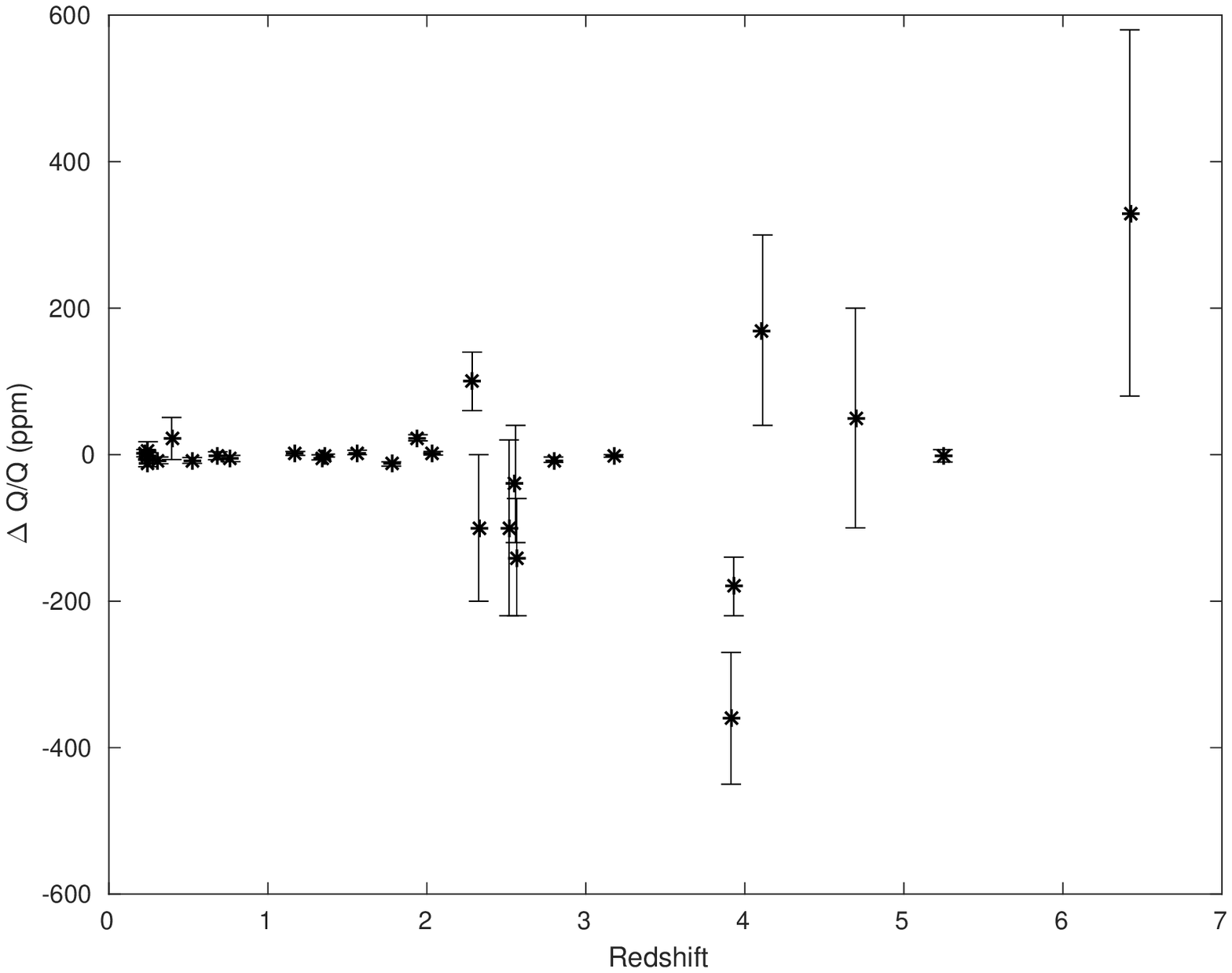}
\end{center}
\caption{\label{fig1}Currently available astrophysical measurements of fundamental couplings: the archival dataset of Webb {\it et al.} (top left), and dedicated measurements of $\alpha$ (top right), $\mu$ (bottom left) and combinations of parameters, generically denoted $Q$ (cf. Table \ref{table1}, bottom right). Note that both the horizontal and the the vertical scales are different in each each panel.}
\end{figure*}

The largest available dataset of $\alpha$ measurements is that of Webb {\it et al.} \cite{Dipole}, containing a total of 293 archival measurements from the HIRES and UVES spectrographs, respectively at the Keck and VLT telescopes. This has been described extensively in the literature, and it is also represented in the top left panel of Fig. \ref{fig1}.

%%%%%%%%%%%%%%%%%%%%%%%%%%%%%%%%%%%%%%%%%%%%%%%%%%%%%%%%%%%%%%%%%%%%%%%%%%%%%%%%%%
%%%%%%%%%%%%%%%%%%%%%%%%%%%%%%%%%%%%%%%%%%%%%%%%%%%%%%%%%%%%%%%%%%%%%%%%%%%%%%%%%%

On the other hand there are various sets of dedicated measurements, which for completeness we will explicitly list in what follows. We will generically present these measurements (and the results of our analysis, in the following sections) in units of parts per million (ppm)---this level of sensitivity is the 'gold standard' for currently available facilities. We typically list (and use in the subsequent analysis) only the tightest available measurement for each astrophysical source. A few older measurements along other lines of sight have not been used, on the grounds that they would have no statistical weight in the analysis. Nevertheless, we will include some low-sensitivity but high-redshift measurements, as these are illustrative of the redshift range that may be probed by future facilities. As in previous work \cite{Ferreira2015}, whose list we update here, our two exceptions regarding measurements of the same source are
\begin{itemize}
\item Measurements using different, independent techniques are both used. Typically this occurs with measurements of $\mu$ or combined measurements using different molecules. Indeed these independent measurements are an important indication of possible systematics.
\item Measurements obtained with different spectrographs. Again, these provide useful clues about possible calibration issues.
\end{itemize}
For these reasons, in these cases we do list the various available measurements. For both the dedicated measurements and the Webb {\it et al.} data, we always consider the total uncertainty, with the statistical and the systematic uncertainty added in quadrature whenever both are available.

\begin{table}
\begin{tabular}{|c|c|c|c|c|}
\hline
Object & z & $Q_{AB}$  & ${ \Delta Q_{AB}}/{Q_{AB}}$ & Ref. \\ 
\hline\hline
J0952$+$179 & 0.234 & ${\alpha^{2}g_{p}/\mu}$ & $2.0\pm5.0$ & \protect\cite{DarlingNew} \\
\hline
PKS1413$+$135 & 0.247 & ${\alpha^{2\times1.85}g_{p}\mu^{1.85}}$  & $-11.8\pm4.6$ & \protect\cite{Kanekar2} \\
PKS1413$+$135 & 0.247 & ${\alpha^{2\times1.57}g_{p}\mu^{1.57}}$  & $5.1\pm12.6$ & \protect\cite{Darling} \\
PKS1413$+$135 & 0.247 & ${\alpha^{2}g_{p}}$  & $-2.0\pm4.4$ & \protect\cite{Murphy} \\
\hline
J1127$-$145 & 0.313 & ${\alpha^{2}g_{p}/\mu}$ & $-7.9\pm4.6$ & \protect\cite{DarlingNew} \\
\hline
J1229$-$021 & 0.395 & ${\alpha^{2}g_{p}/\mu}$ & $20.1\pm28.7$ & \protect\cite{DarlingNew} \\
\hline
J0235$+$164 & 0.524 & ${\alpha^{2}g_{p}/\mu}$ & $-8.0\pm3.9$ & \protect\cite{DarlingNew} \\
\hline
B0218$+$357 & 0.685 & ${\alpha^{2}g_{p}}$ & $-1.6\pm5.4$ & \protect\cite{Murphy} \\
\hline
J0134$-$0931 & 0.765 & ${\alpha^{2\times1.57}g_{p}\mu^{1.57}}$  &  $-5.2\pm4.3$ & \protect\cite{Kanekar} \\
\hline
J2358$-$1020 & 1.173 & ${\alpha^{2}g_{p}/\mu}$ & $1.8\pm2.7$ & \protect\cite{Rahmani} \\
\hline
J1623$+$0718 & 1.336 & ${\alpha^{2}g_{p}/\mu}$ & $-3.7\pm3.4$ & \protect\cite{Rahmani} \\
\hline
J2340$-$0053 & 1.361 & ${\alpha^{2}g_{p}/\mu}$ & $-1.3\pm2.0$ & \protect\cite{Rahmani} \\
\hline
J0501$-$0159 & 1.561 & ${\alpha^{2}g_{p}/\mu}$ & $3.0\pm3.1$ & \protect\cite{Rahmani} \\
\hline
J1381$+$170 & 1.776 & ${\alpha^{2}g_{p}/\mu}$ & $-12.7\pm3.0$ & \protect\cite{DarlingNew} \\
\hline
J1157$+$014 & 1.944 & ${\alpha^{2}g_{p}/\mu}$ & $23.1\pm4.2$ & \protect\cite{DarlingNew} \\
\hline
J0458$-$020 & 2.040 & ${\alpha^{2}g_{p}/\mu}$ & $1.9\pm2.5$ & \protect\cite{DarlingNew} \\
\hline
J1024$+$4709 & 2.285 & ${\alpha^{2}\mu}$ & $100\pm40$ & \protect\cite{Curran} \\
\hline
J2135$-$0102 & 2.326 & ${\alpha^{2}\mu}$ & $-100\pm100$ & \protect\cite{Curran} \\
\hline
J1636$+$6612 & 2.517 & ${\alpha^{2}\mu}$ & $-100\pm120$ & \protect\cite{Curran} \\
\hline
H1413$+$117 & 2.558 & ${\alpha^{2}\mu}$ & $-40\pm80$ & \protect\cite{Curran} \\
\hline
J1401$+$0252 & 2.565 & ${\alpha^{2}\mu}$ & $-140\pm80$ & \protect\cite{Curran} \\
\hline
J0911$+$0551 & 2.796 & ${\alpha^{2}\mu}$ & $ -6.9\pm3.7$ & \protect\cite{Weiss} \\
\hline
J1337$+$3152  & 3.174 & ${\alpha^{2}g_{p}/\mu}$ & $-1.7\pm1.7$ & \protect\cite{Petitjean1} \\
\hline
APM0828$+$5255 & 3.913 & ${\alpha^{2}\mu}$ & $-360\pm90$ & \protect\cite{Curran} \\
\hline
MM1842$+$5938 & 3.930 & ${\alpha^{2}\mu}$ & $-180\pm40$ & \protect\cite{Curran} \\
\hline
PSS2322$+$1944 & 4.112 & ${\alpha^{2}\mu}$ & $170\pm130$ & \protect\cite{Curran} \\
\hline
BR1202$-$0725 & 4.695 & ${\alpha^{2}\mu}$ & $50\pm150$ & \protect\cite{Lentati} \\
\hline
J0918$+$5142 & 5.245 & ${\alpha^{2}\mu}$ & $-1.7\pm8.5$ & \protect\cite{Levshakov} \\
\hline
J1148$+$5251 & 6.420 & ${\alpha^{2}\mu}$ & $330\pm250$ & \protect\cite{Lentati} \\
\hline
\end{tabular}
\caption{\label{table1}Available measurements of several combinations of the dimensionless couplings $\alpha$, $\mu$ and $g_p$. Listed are, respectively, the object along each line of sight, the redshift of the measurement, the dimensionless parameter being constrained, the measurement itself (in parts per million), and its original reference.}
\end{table}

Table \ref{table1} and the bottom right panel of Fig. \ref{fig1} contain current joint measurements of several combinations of couplings. Compared to our earlier work we have added the 7 measurements of \cite{DarlingNew}, whose individual values were kindly provided by the author. Note that for the radio source PKS1413$+$135 the three available measurements are sufficient to yield individual constraints on the variations of the three quantities at redshift $z=0.247$. This analysis was done in \cite{PKS}, yielding a null result at the two sigma confidence level.

%%%%%%%%%%%%%%%%%%%%%%%%%%%%%%%%%%%%%%%%%%%%%%%%%%%%%%%%%%%%%%%%%%%%%%%%%%%%%%%%%%
\begin{table}
\begin{center}
\begin{tabular}{|c|c|c|c|c|}
\hline
 Object & z & ${ \Delta\alpha}/{\alpha}$ (ppm) & Spectrograph & Ref. \\
\hline\hline
J0026$-$2857 & 1.02 & $3.5\pm8.9$ & UVES & \protect\cite{MalecNew} \\
\hline
J0058$+$0041 & 1.07 & $-1.4\pm7.2$ & HIRES & \protect\cite{MalecNew} \\
\hline
3 sources & 1.08 & $4.3\pm3.4$ & HIRES & \protect\cite{Songaila} \\
\hline
HS1549$+$1919 & 1.14 & $-7.5\pm5.5$ & UVES/HIRES/HDS & \protect\cite{LP3} \\
\hline
HE0515$-$4414 & 1.15 & $-1.4\pm0.9$ & UVES & \protect\cite{Kotus} \\
\hline
J1237$+$0106 & 1.31 & $-4.5\pm8.7$ & HIRES & \protect\cite{MalecNew} \\
\hline
HS1549$+$1919 & 1.34 & $-0.7\pm6.6$ & UVES/HIRES/HDS & \protect\cite{LP3} \\
\hline
J0841$+$0312 & 1.34 & $3.0\pm4.0$ & HIRES & \protect\cite{MalecNew} \\
J0841$+$0312 & 1.34 & $5.7\pm4.7$ & UVES & \protect\cite{MalecNew} \\
\hline
J0108$-$0037 & 1.37 & $-8.4\pm7.3$ & UVES & \protect\cite{MalecNew} \\
\hline
HE0001$-$2340 & 1.58 & $-1.5\pm2.6$ &  UVES & \protect\cite{alphaAgafonova}\\
\hline
J1029$+$1039 & 1.62 & $-1.7\pm10.1$ & HIRES & \protect\cite{MalecNew} \\
\hline
HE1104$-$1805 & 1.66 & $-4.7\pm5.3$ & HIRES & \protect\cite{Songaila} \\
\hline
HE2217$-$2818 & 1.69 & $1.3\pm2.6$ &  UVES & \protect\cite{LP1}\\
\hline
HS1946$+$7658 & 1.74 & $-7.9\pm6.2$ & HIRES & \protect\cite{Songaila} \\
\hline
HS1549$+$1919 & 1.80 & $-6.4\pm7.2$ & UVES/HIRES/HDS & \protect\cite{LP3} \\
\hline
Q1103$-$2645 & 1.84 & $3.5\pm2.5$ &  UVES & \protect\cite{Bainbridge}\\
\hline
Q2206$-$1958 & 1.92 & $-4.6\pm6.4$ &  UVES & \protect\cite{MalecNew}\\
\hline
Q1755$+$57 & 1.97 & $4.7\pm4.7$ & HIRES & \protect\cite{MalecNew} \\
\hline
PHL957 & 2.31 & $-0.7\pm6.8$ & HIRES & \protect\cite{MalecNew} \\
PHL957 & 2.31 & $-0.2\pm12.9$ & UVES & \protect\cite{MalecNew} \\
\hline
\end{tabular}
\caption{\label{table2}Available dedicated measurements of $\alpha$. Listed are, respectively, the object along each line of sight, the redshift of the measurement, the measurement itself (in parts per million), the spectrograph, and the original reference. The third measurement is the weighted average from 8 absorbers along the lines of sight of HE1104-1805A, HS1700+6416 and HS1946+7658, reported in \cite{Songaila} without the values for individual systems.}
\end{center}
\end{table}
%%%%%%%%%%%%%%%%%%%%%%%%%%%%%%%%%%%%%%%%%%%%%%%%%%%%%%%%%%%%%%%%%%%%%%%%%%%%%%%%%%

Table \ref{table2} contains the individual dedicated $\alpha$ measurements, which are also depicted in the top right panel of Fig. \ref{fig1}. Compared to our previous analysis, there are 11 new measurements from \cite{MalecNew}, as well as improved measurements of two previously observed targets \cite{Bainbridge,Kotus}. The latter of these is currently the tightest individual measurement of $\alpha$. We note that the weighted mean of the measurements on the table is
\begin{equation}\label{prioralpha}
\left(\frac{\Delta\alpha}{\alpha}\right)_{wm}=-0.64\pm0.65\, ppm\,,
\end{equation}
and thus compatible with the null result, unlike the archival dataset of Webb {\it et al.}, for which the weighted mean is nominally \cite{KingDip}
\begin{equation}\label{priorwebb}
\left(\frac{\Delta\alpha}{\alpha}\right)_{wm}=-2.16\pm0.86\, ppm\,.
\end{equation}

\begin{table}
\begin{tabular}{|c|c|c|c|c|}
\hline
 Object & z & ${\Delta\mu}/{\mu}$ & Method & Ref. \\ 
\hline\hline
B0218$+$357 & 0.685 & $0.74\pm0.89$ & $NH_3$/$HCO^+$/$HCN$ & \protect\cite{Murphy2} \\
B0218$+$357 & 0.685 & $-0.35\pm0.12$ & $NH_3$/$CS$/$H_2CO$ & \protect\cite{Kanekar3} \\
\hline
PKS1830$-$211 & 0.886 & $0.08\pm0.47$ &  $NH_3$/$HC_3N$ & \protect\cite{Henkel}\\
PKS1830$-$211 & 0.886 & $-1.2\pm4.5$ &  $CH_3NH_2$ & \protect\cite{Ilyushin}\\
PKS1830$-$211 & 0.886 & $-2.04\pm0.74$ & $NH_3$ & \protect\cite{Muller}\\
PKS1830$-$211 & 0.886 & $-0.10\pm0.13$ &  $CH_3OH$ & \protect\cite{Bagdonaite2}\\
\hline
J2123$-$005 & 2.059 & $8.5\pm4.2$ & $H_2$/$HD$ (VLT)& \protect\cite{vanWeerd} \\
J2123$-$005 & 2.059 & $5.6\pm6.2$ & $H_2$/$HD$ (Keck)& \protect\cite{Malec} \\
\hline
HE0027$-$1836 & 2.402 & $-7.6\pm10.2$ & $H_2$ & \protect\cite{LP2} \\
\hline
Q2348$-$011 & 2.426 & $-6.8\pm27.8$ & $H_2$ & \protect\cite{Bagdonaite} \\
\hline
Q0405$-$443 & 2.597 & $10.1\pm6.2$ & $H_2$ & \protect\cite{King} \\
\hline
J0643$-$504 & 2.659 & $7.4\pm6.7$ & $H_2$ & \protect\cite{Albornoz} \\
\hline
J1237$+$0648 & 2.688 & $-5.4\pm7.5$ & $H_2$/$HD$ & \protect\cite{Dapra} \\
\hline
Q0528$-$250 & 2.811 & $0.3\pm3.7$ & $H_2$/$HD$ & \protect\cite{King2} \\
\hline
Q0347$-$383 & 3.025 & $2.1\pm6.0$ & $H_2$ & \protect\cite{Wendt} \\
\hline
J1443$+$2724 & 4.224 & $-9.5\pm7.6$ & $H_2$ & \protect\cite{Bagdonaite4} \\
\hline
\end{tabular}
\caption{\label{table3}Available measurements of $\mu$. Listed are, respectively, the object along each line of sight, the redshift of the measurement, the measurement itself, the molecule(s) used, and the original reference.}
\end{table}

Table \ref{table3} contains individual $\mu$ measurements, which are shown in the bottom left panel of Fig. \ref{fig1}. Note that several different molecules can be used, and in the case of the gravitational lens PKS1830$-$211 there are actually four independent measurements, with different levels of sensitivity. Currently ammonia is the most common molecule at low redshift, though others such as methanol, peroxide, hydronium and methanetiol have a greater potential in the context of facilities like ALMA \cite{Molecules}. At higher redshifts molecular hydrogen is the most common.

The tightest available constraint on $\mu$ comes precisely from PKS1830$-$211, from observations of methanol transitions \cite{Bagdonaite2}. With respect to our previous compilation there is one new measurement, from \cite{Dapra}. We can similarly calculate the weighted mean of the low and high-redshift samples ($z<1$ and $z>2$ respectively), we find
\begin{equation}\label{priormulo}
\left(\frac{\Delta\mu}{\mu}\right)_{Low,wm}=-0.24\pm0.09\, ppm\,
\end{equation}
\begin{equation}\label{priormuhi}
\left(\frac{\Delta\mu}{\mu}\right)_{High,wm}=2.9\pm1.9\, ppm\,,
\end{equation}
in both cases this is weak evidence for a variation, although note the preferred sign of this variation is different at high and low redshifts.

%%%%%%%%%%%%%%%%%%%%%%%%%%%%%%%%%%%%%%%%%%%%%%%%%%%%%%%%%%%%%%%%%%%%%%%%%%%%%%
\section{\label{bestfit}Consistency between datasets}

We start by analyzing the combined measurements of Table I. These constrain combinations of the fine-structure constant, the proton-to-electron mass ratio and the proton gyromagnetic ratio, which we can generically write
\begin{equation}\label{defq}
Q(a,b,c)=\alpha^a\mu^bg_p^c\,;
\end{equation}
the exponents are different for each measurement, as explicitly indicated for each measurement in the table itself. We can therefore write
\begin{equation}\label{varq}
\frac{\Delta Q}{Q}=a \frac{\Delta \alpha}{\alpha}+b\frac{\Delta\mu}{\mu}+c\frac{\Delta g_p}{g_p}\,,
\end{equation}
and can straightforwardly look for the best-fit values of the relative variations of each of the individual couplings. In this section we will assume that there is a single such value, regardless of the redshift of the measurement; we will relax this assumption in the following section.

\begin{figure}
\begin{center}
\includegraphics[width=2.6in]{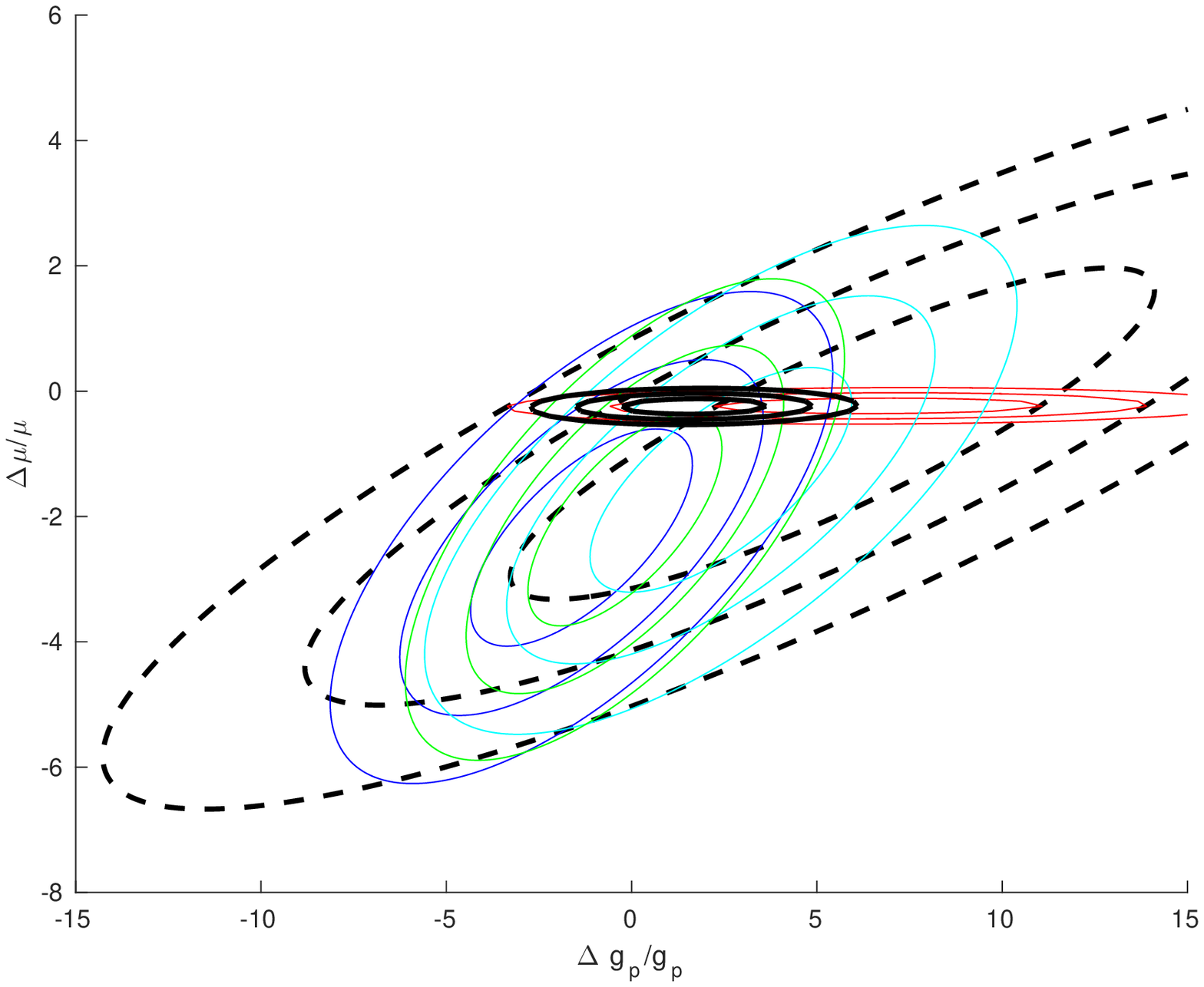}
\vskip0.25in
\includegraphics[width=2.6in]{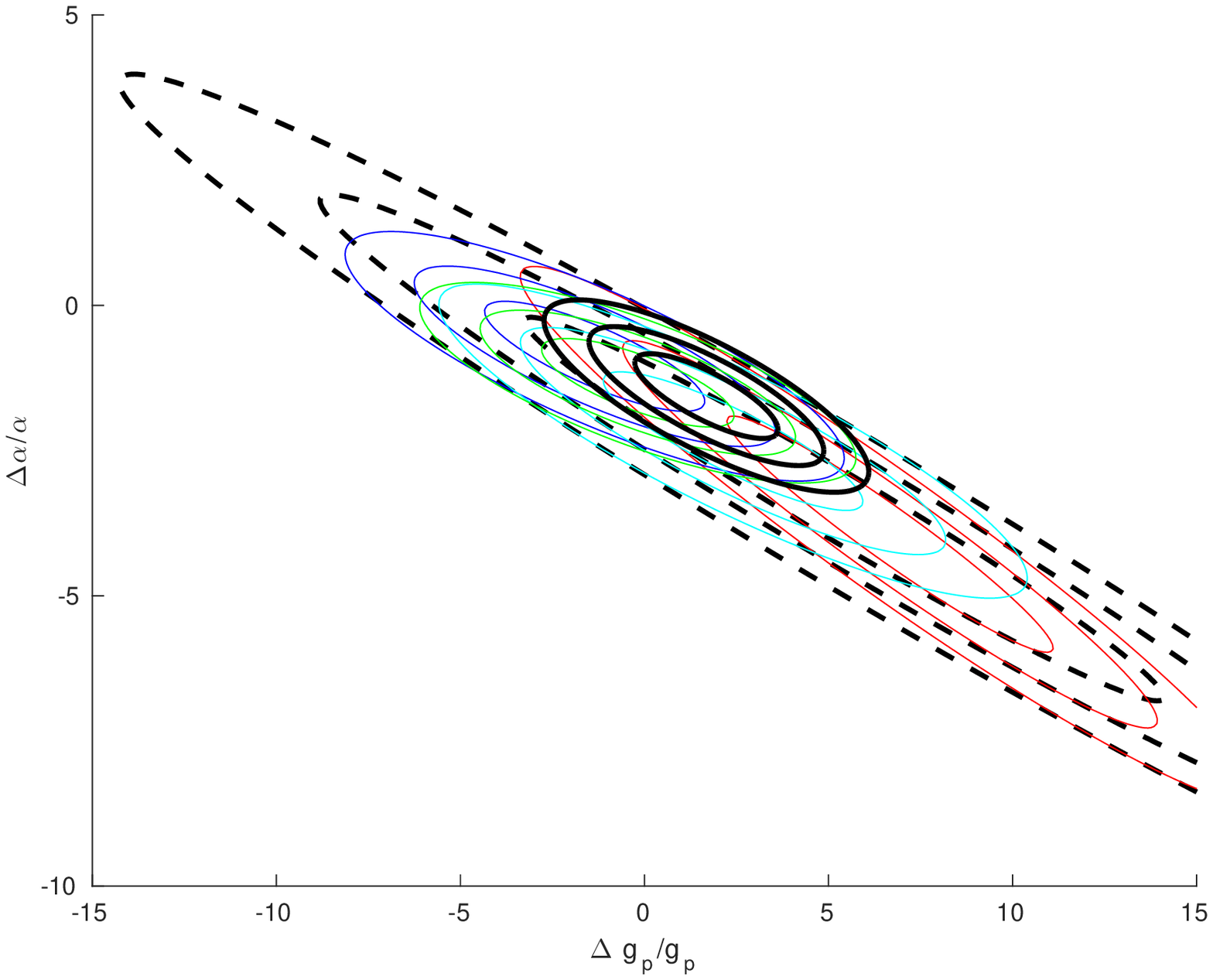}
\vskip0.25in
\includegraphics[width=2.6in]{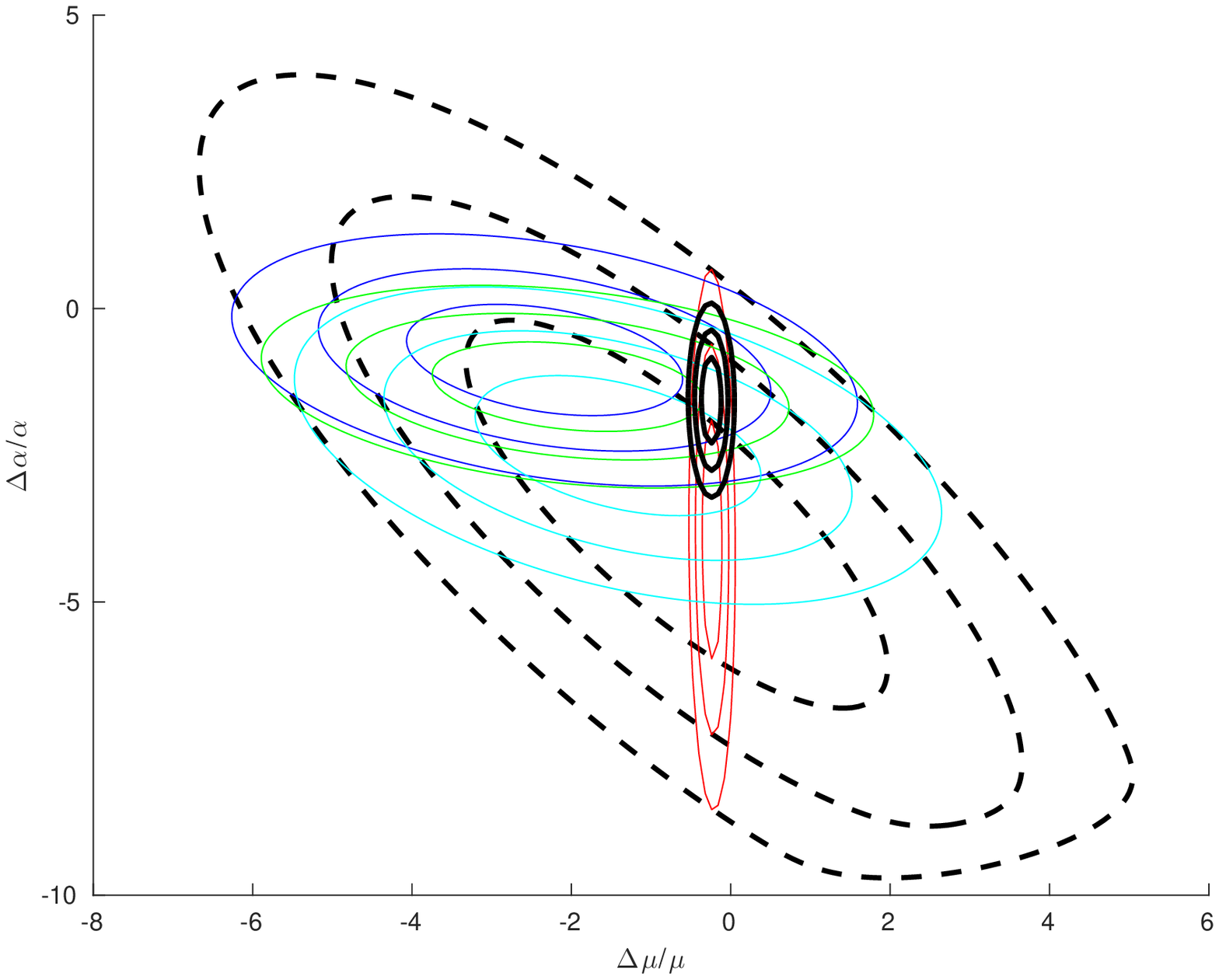}
\end{center}
\caption{\label{fig2}2D likelihoods (marginalizing over the third parameter) for the three dimensionless couplings and various combinations of datasets: Table I (combined measurements) is shown with black dashed lines, Table I plus Table II (dedicated $\alpha$ measurements) in blue, Table I plus Webb {\it et al.}in cyan, Table I plus Table II plus Webb {\it et al.} (in other words, all direct $\alpha$ measurements) in green, Table I plus Table III (dedicated $\mu$ measurements) in red, and the full dataset in solid black. One, two and three sigma confidence levels are shown in all cases.}
\end{figure}

\begin{figure}
\begin{center}
\includegraphics[width=2.6in]{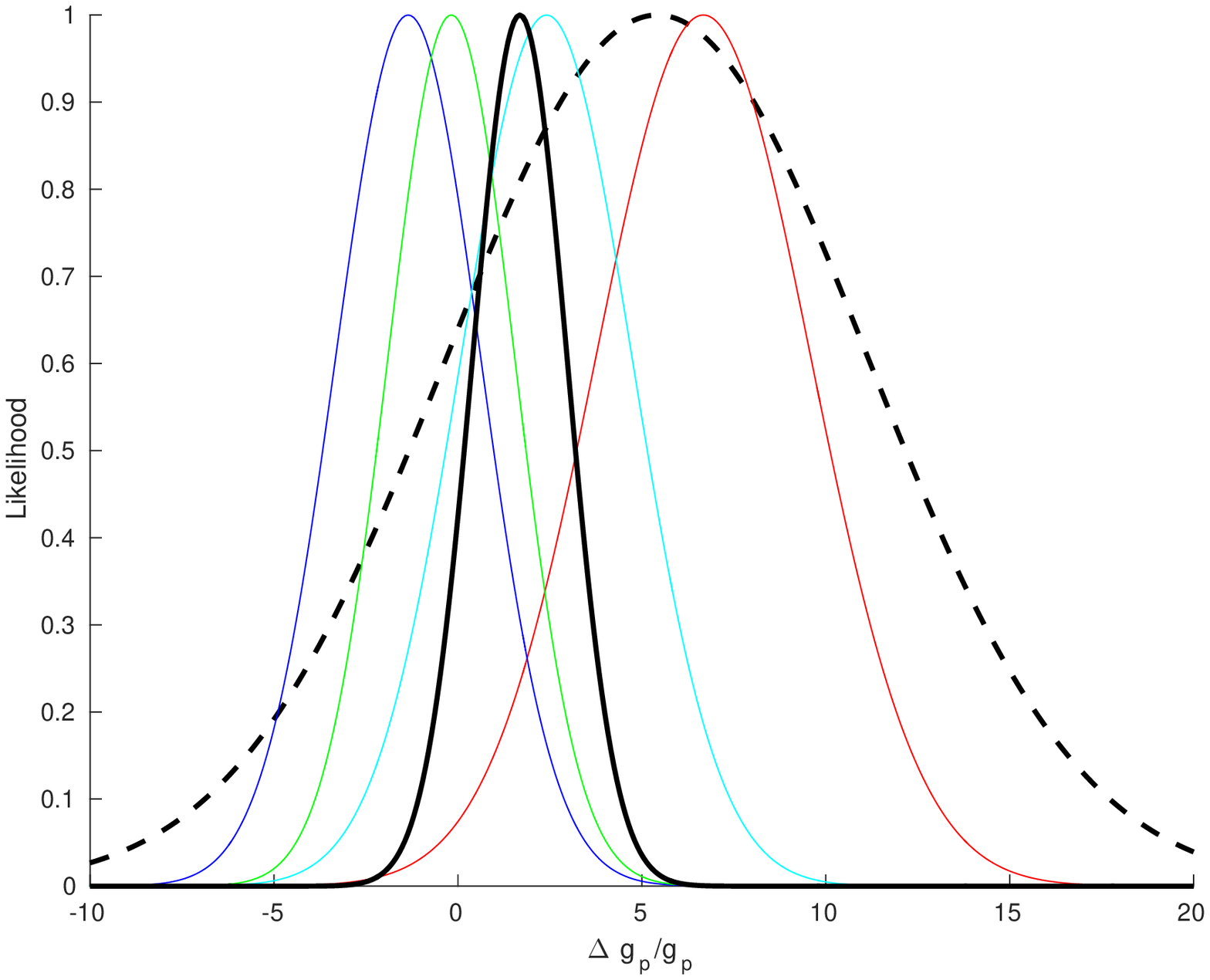}
\vskip0.25in
\includegraphics[width=2.6in]{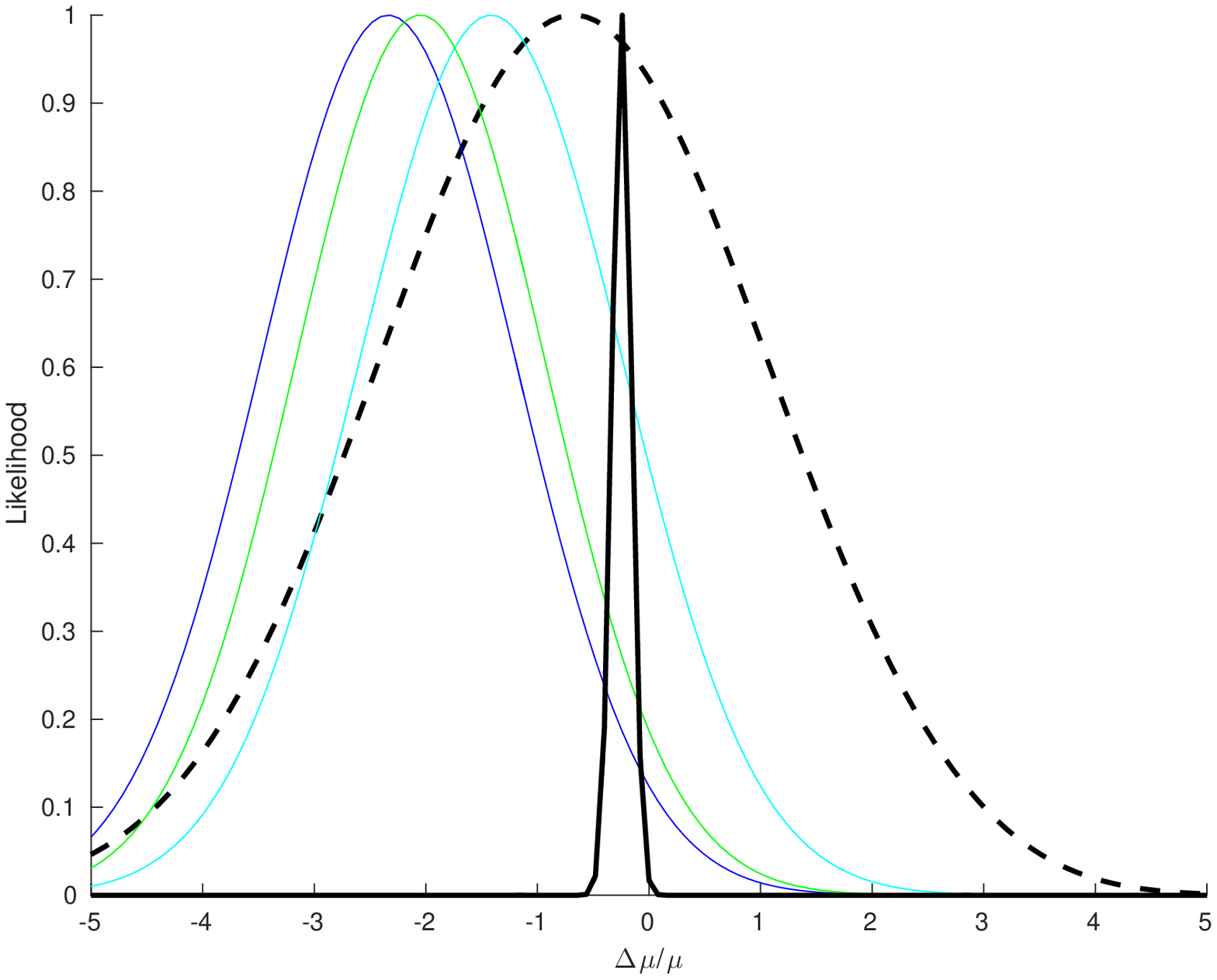}
\vskip0.25in
\includegraphics[width=2.6in]{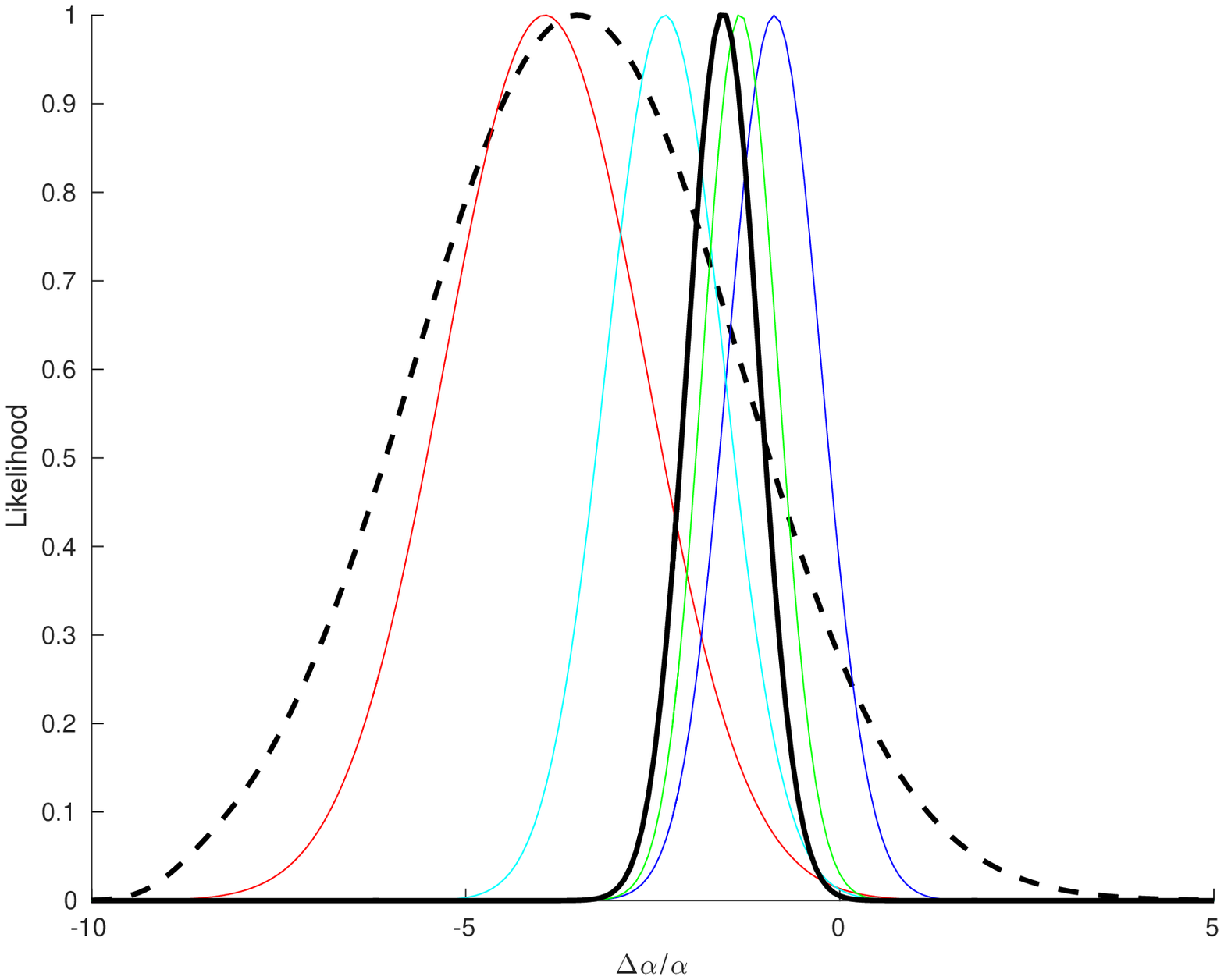}
\end{center}
\caption{\label{fig3}1D likelihoods (marginalizing over the other parameter) for $\alpha$, $\mu$ and $g_p$, for the same combinations of datasets (and the same color conventions) described in Figure \protect\ref{fig2}.}
\end{figure}

On the other hand, we can also combine the measurements of Table I with the direct measurements of $\alpha$, $\mu$ or both (those of Webb {\it et al.} as well as those listed in Tables II and III); these will effectively provide priors on the respective parameters, leading to stronger constraints. The results of this analysis are summarized in Figs. \ref{fig2} and \ref{fig3}, as well as in Table \ref{table4}

%%%%%%%%%%%%%%%%%%%%%%%%%%%%%%%%%%%%%%%%%%%%%%%%%%%%%%%%%%%%%%%%%%%%%%%%%%%%%%%%%%
%%%%%%%%%%%%%%%%%%%%%%%%%%%%%%%%%%%%%%%%%%%%%%%%%%%%%%%%%%%%%%%%%%%%%%%%%%%%%%%%%%
%%%%%%%%%%%%%%%%%%%%%%%%%%%%%%%%%%%%%%%%%%%%%%%%%%%%%%%%%%%%%%%%%%%%%%%%%%%%%%%%%%
\begin{table*}
\begin{center}
\begin{tabular}{|c|c|c|c|c|}
\hline
 Sample & ${ \Delta\alpha}/{\alpha}$ (ppm) & ${ \Delta\mu}/{\mu}$ (ppm) & ${\Delta g_p}/{g_p}$ (ppm) & $\chi^2_\nu$  \\
\hline
Table I only &  $-3.5\pm2.2$ & $-0.6\pm1.7$ & $5.4\pm5.7$ & 3.83 \\
\hline
Table I + Webb &  $-2.3\pm0.8$ &  $-1.4\pm1.2$ & $2.4\pm2.4$ & 1.28 \\
Table I + II &  $-0.9\pm0.6$ & $-2.3\pm1.1$ & $-1.4\pm2.0$ & 2.58 \\
Table I + II + Webb &  $-1.4\pm0.5$ &  $-2.1\pm1.1$ & $-0.2\pm1.7$ & 1.26 \\
\hline
Table I + III & $-3.9\pm1.3$ & $-0.2\pm0.1$ & $6.6\pm2.9$ & 2.95 \\
\hline
All data & $-1.6\pm0.5$ & $-0.2\pm0.1$ & $1.7\pm1.3$ & 1.27 \\
\hline
\end{tabular}
\caption{\label{table4}One-dimensional marginalized one-sigma constraints for $\alpha$, $\mu$ an $g_p$, for various combinations of datasets. All constraints are in parts per million. The last column has the reduced chi-square for maximum of the 3D likelihood.}
\end{center}
\end{table*}
%%%%%%%%%%%%%%%%%%%%%%%%%%%%%%%%%%%%%%%%%%%%%%%%%%%%%%%%%%%%%%%%%%%%%%%%%%%%%%%%%%

At face value there is a mild preference, at the level of two to three standard deviations, for negative variations of $\alpha$ and $\mu$. However, the most noteworthy result of this analysis are the very large values of the reduced chi-square at the maximum of the three-dimensional likelihoods, which are also listed in Table \ref{table4}. This is mostly due to the combined measurements dataset, but the issue remains when Table I is combined with Table II or Table III. One possible explanation is that the uncertainties of some of the measurements have been underestimated. However, this also suggests that assuming a single redshift-independent value for each parameter may not be an adequate assumption: indeed, for Table III this is manifest in the fact that the weighted mean values of the low and high redshift subsamples are clearly different. Our next step is therefore to repeat this analysis by dividing our data into several redshift bins.

%%%%%%%%%%%%%%%%%%%%%%%%%%%%%%%%%%%%%%%%%%%%%%%%%%%%%%%%%%%%%%%%%%%%%%%%%%%%%%
\section{\label{zbins}Redshift tomography}

We will now divide the data into four different redshift bins. The first three have bin size $\Delta z=1$, while the fourth is includes the data with $z>3$. This is a convenient statistical division, but it also makes sense from an observational point of view. This is clearest for direct $\mu$ measurements, as can be seen in Table III: in the radio/mm they are all at low redshifts ($z<1$) while those in the optical using molecular hydrogen are at $z>2$. Similarly, for direct $\alpha$ measurements, different atomic transitions are typically used at high and low redshifts; here the transition redshift is not sharp, but it is around $z\sim2$. Finally, it is also clear from Table I that different combinations of the couplings can be measured at different redshifts.

%%%%%%%%%%%%%%%%%%%%%%%%%%%%%%%%%%%%%%%%%%%%%%%%%%%%%%%%%%%%%%%%%%%%%%%%%%%%%%%%%%
\begin{figure}
\begin{center}
\includegraphics[width=2.6in]{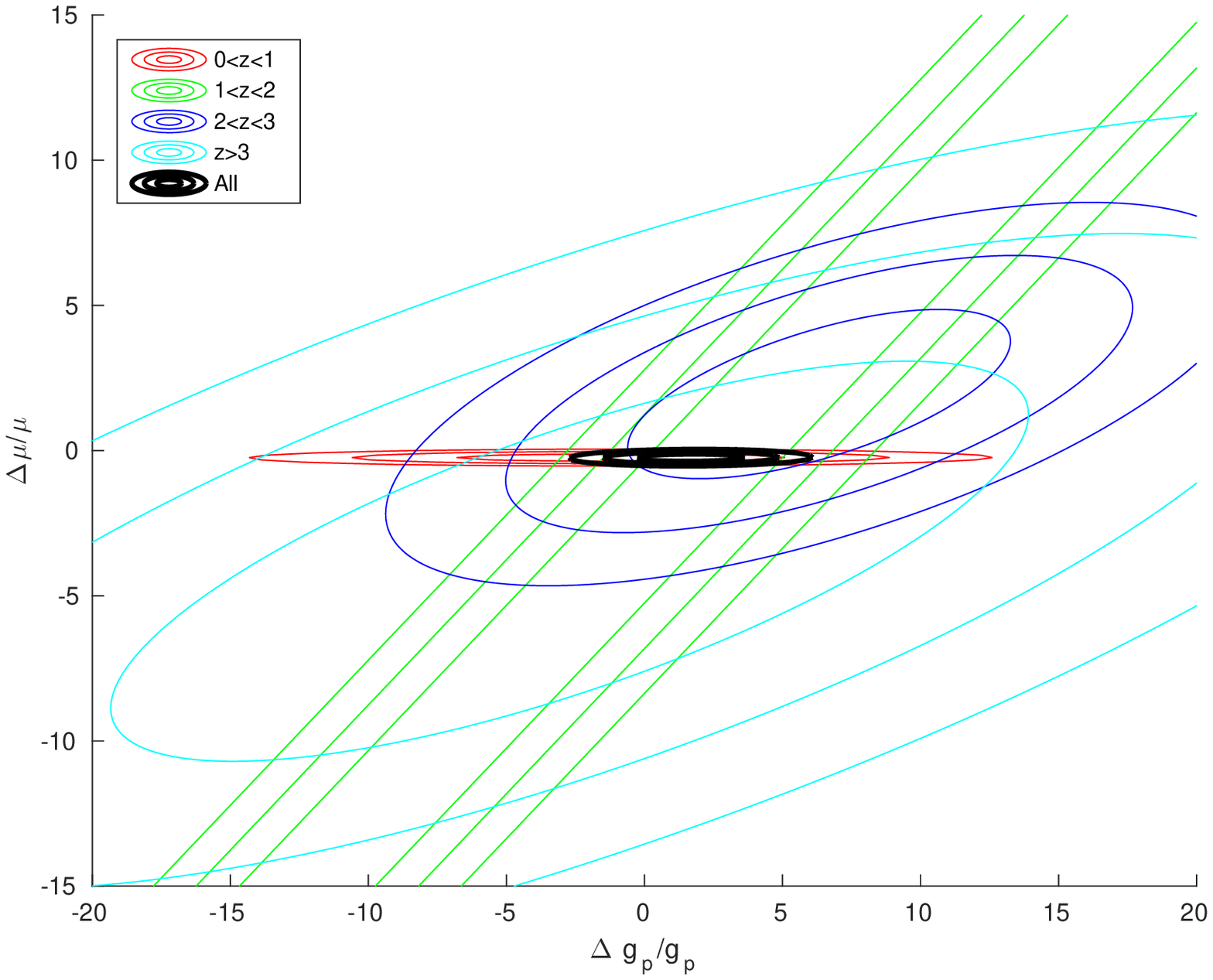}
\vskip0.25in
\includegraphics[width=2.6in]{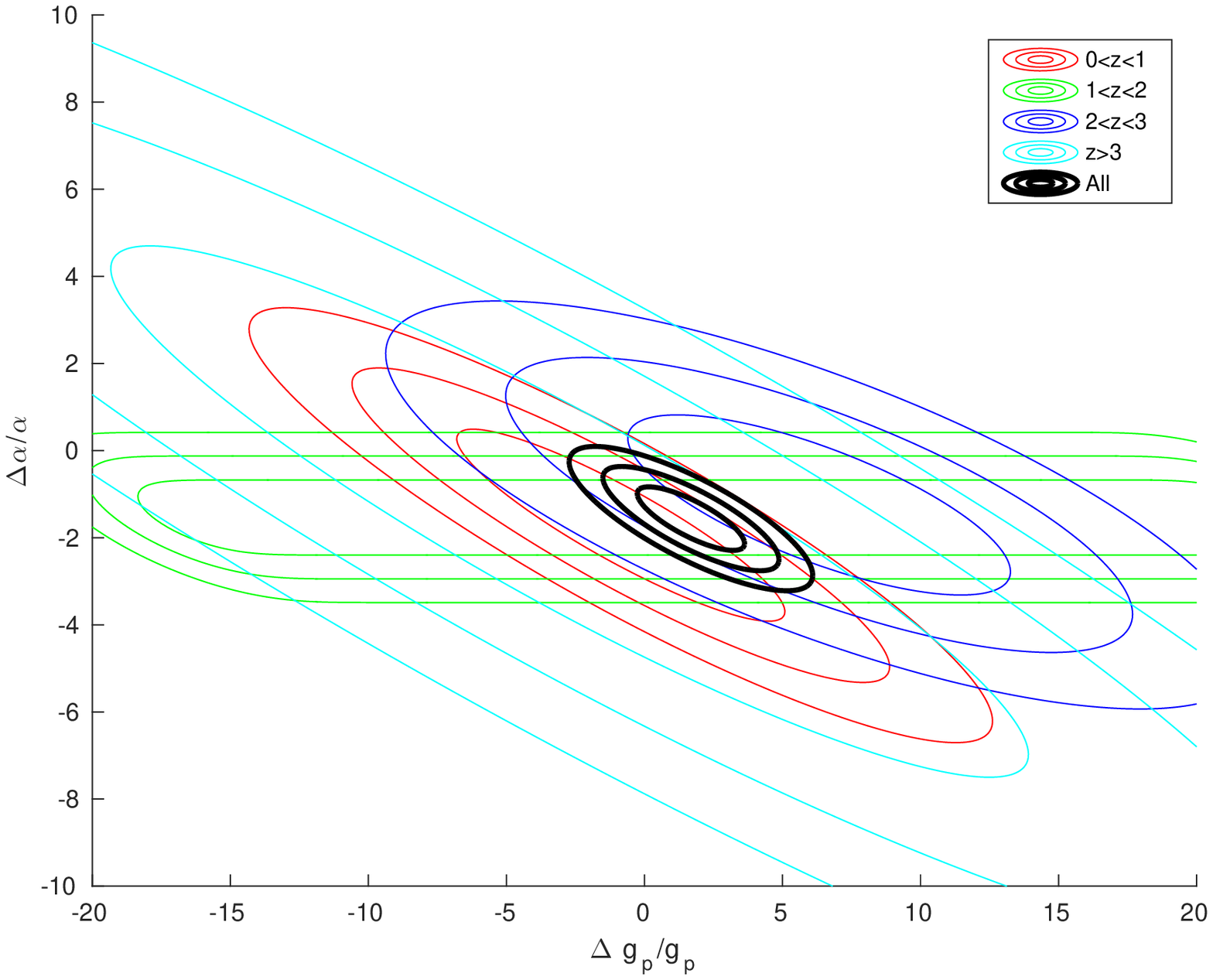}
\vskip0.25in
\includegraphics[width=2.6in]{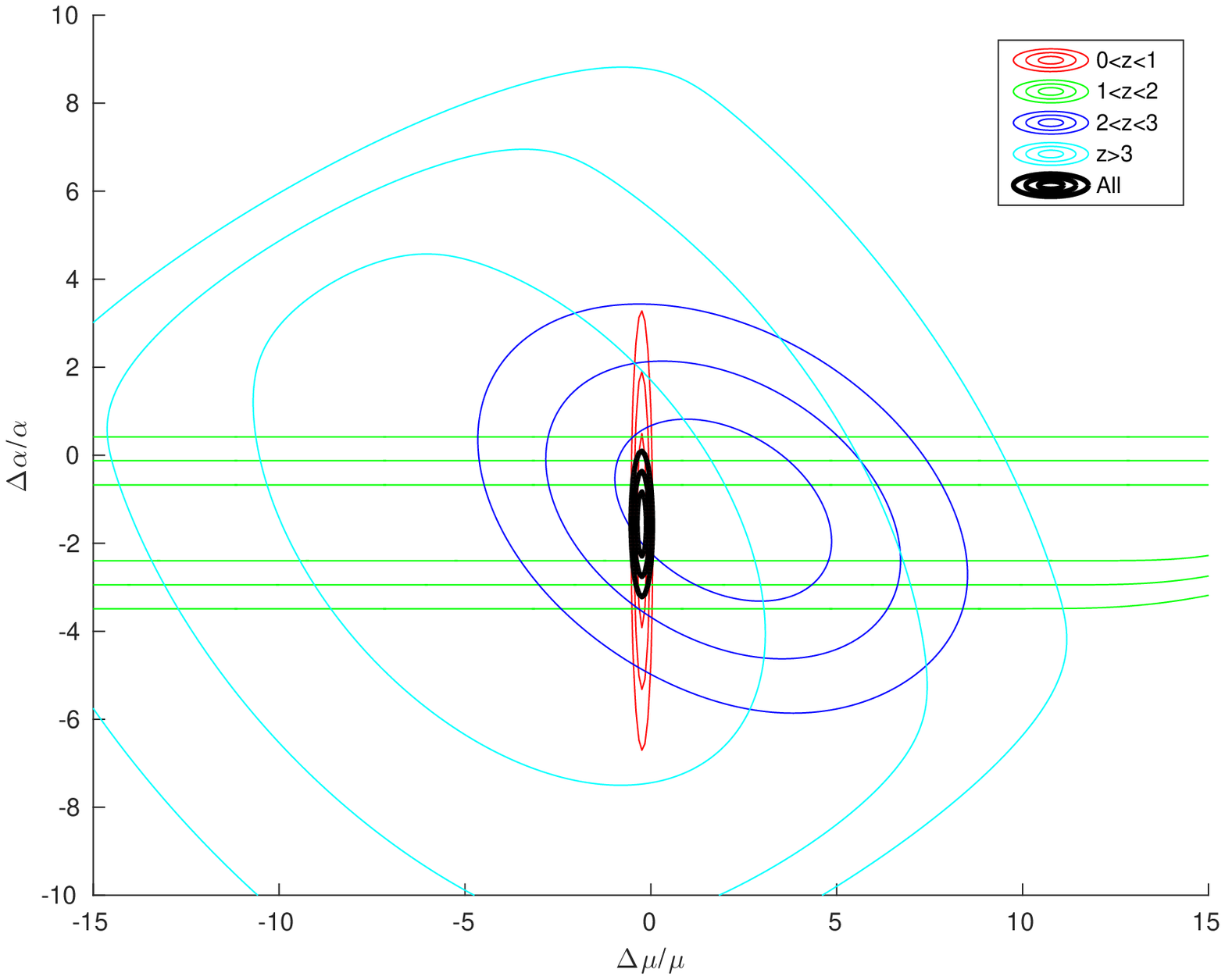}
\end{center}
\caption{\label{fig4}2D likelihoods (marginalizing over the third parameter) for the three dimensionless couplings in four different redshift bins (red, green, blue and cyan contours), as well as for the full dataset (black contours). One, two and three sigma contours are shown in all cases.}
\end{figure}
%%%%%%%%%%%%%%%%%%%%%%%%%%%%%%%%%%%%%%%%%%%%%%%%%%%%%%%%%%%%%%%%%%%%%%%%%%%%%%%%%%
\begin{figure}
\begin{center}
\includegraphics[width=2.6in]{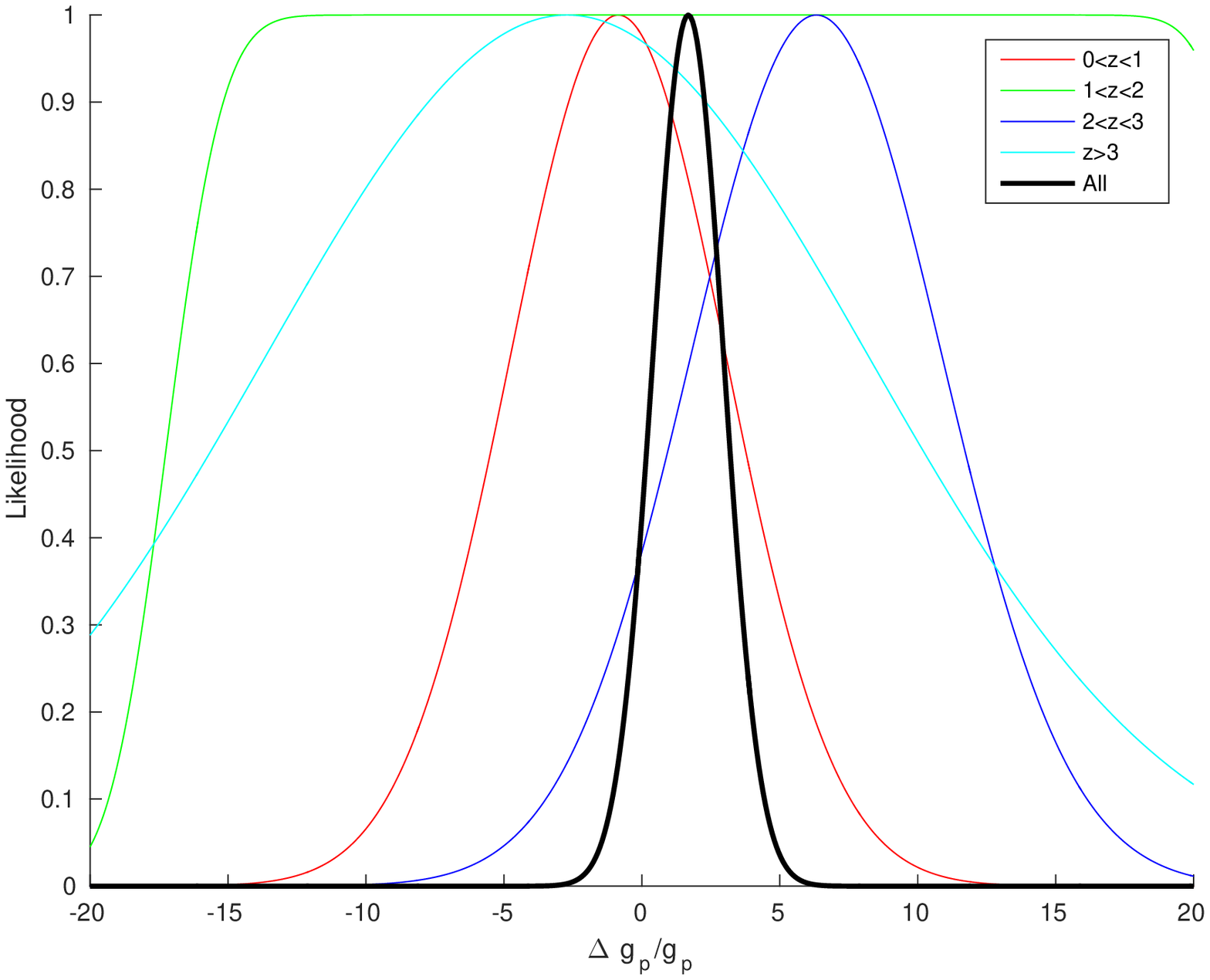}
\vskip0.25in
\includegraphics[width=2.6in]{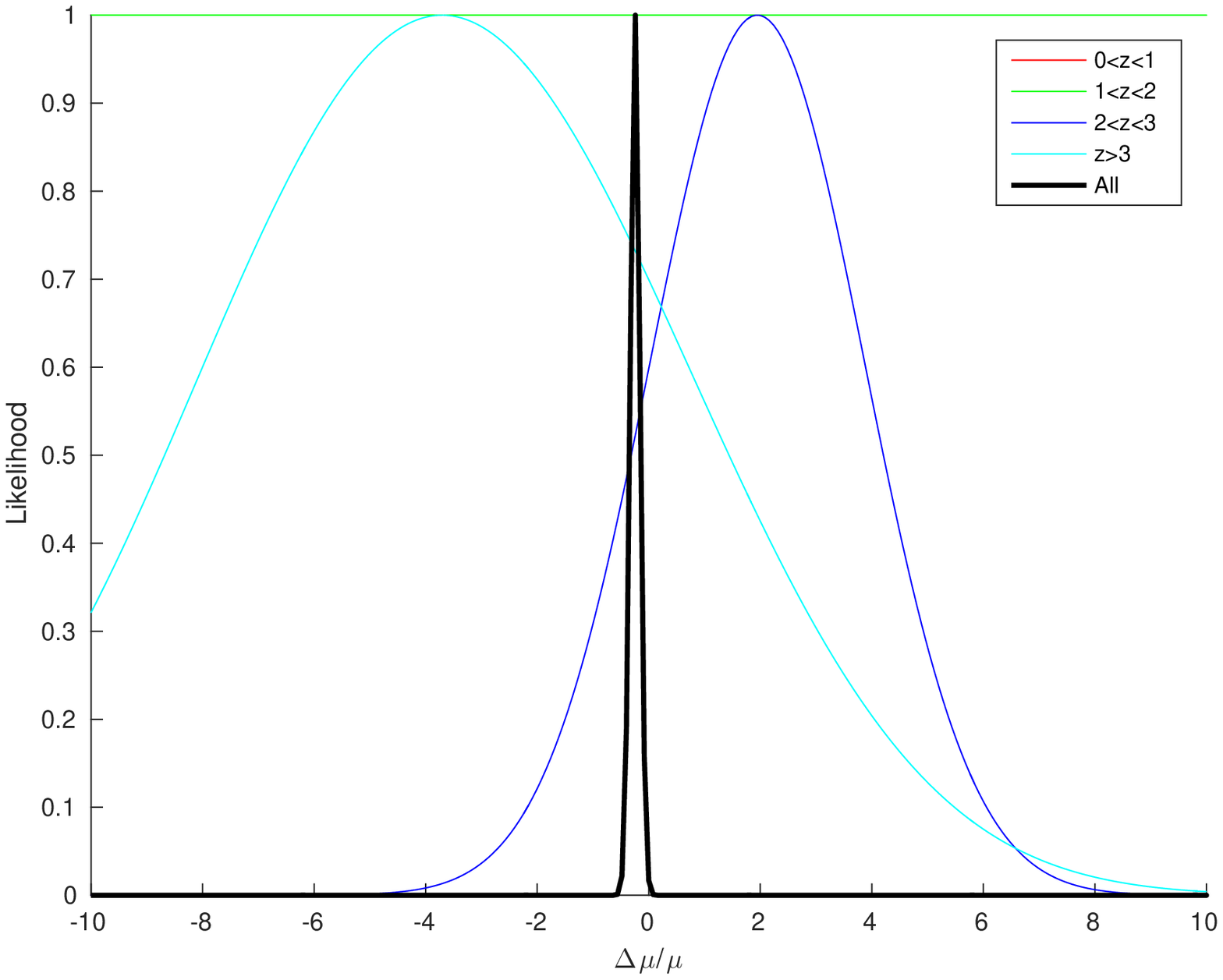}
\vskip0.25in
\includegraphics[width=2.6in]{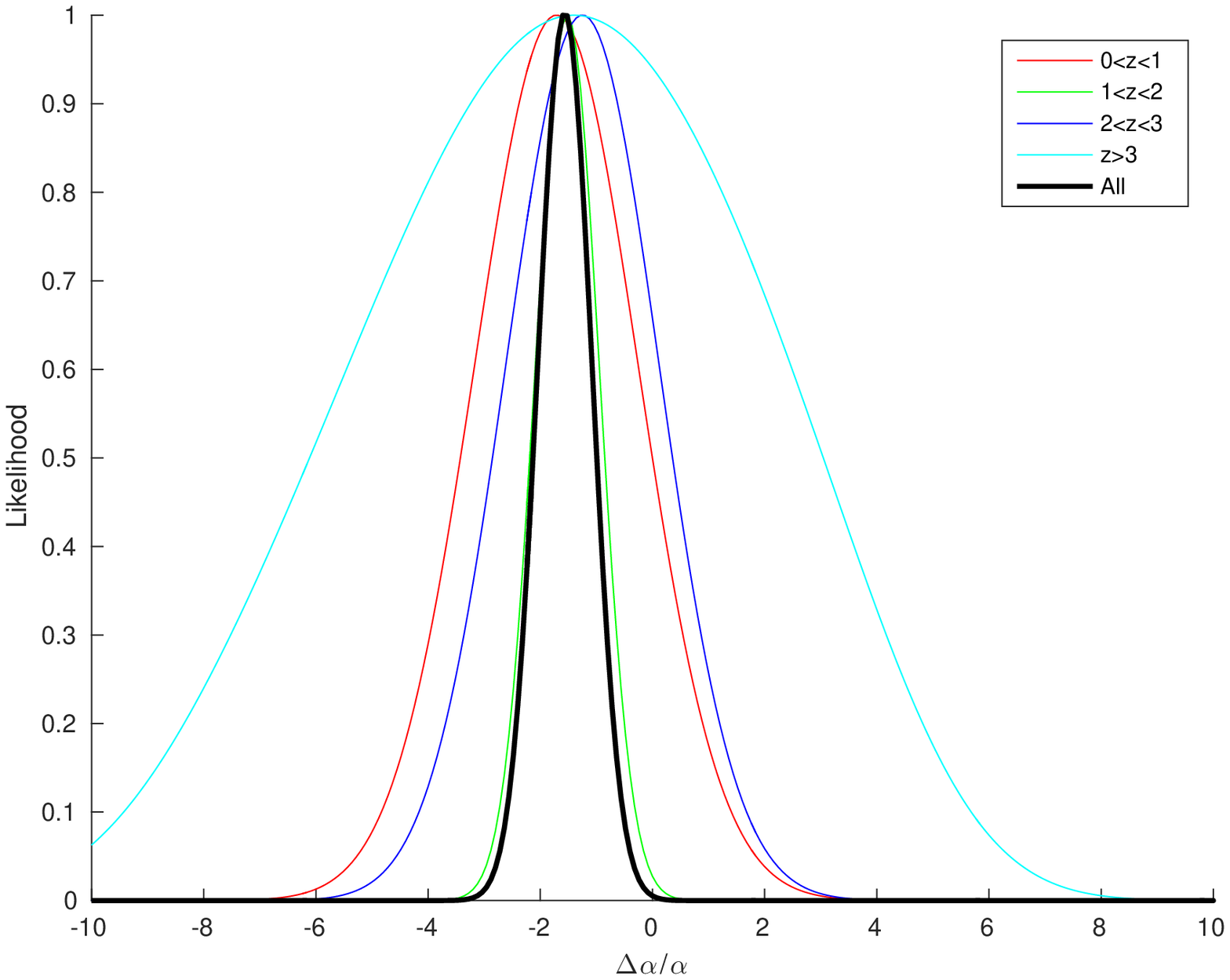}
\end{center}
\caption{\label{fig5}1D likelihoods (marginalizing over the other parameter) for $\alpha$, $\mu$ and $g_p$, for the same redshift bins of Figure \protect\ref{fig4}.}
\end{figure}
%%%%%%%%%%%%%%%%%%%%%%%%%%%%%%%%%%%%%%%%%%%%%%%%%%%%%%%%%%%%%%%%%%%%%%%%%%%%%%%%%%
\begin{table*}
\begin{center}
\begin{tabular}{|c|ccccc|c|c|c|c|}
\hline
Redshift & Tab.I & Webb & Tab.II & Tab.III & Total & ${ \Delta\alpha}/{\alpha}$ (ppm) & ${ \Delta\mu}/{\mu}$ (ppm) & ${\Delta g_p}/{g_p}$ (ppm) & $\chi^2_\nu$  \\
\hline
$0<z<1$ & 9 & 64 & 0 & 6 & 79 &  $-1.7\pm1.5$ & $-0.2\pm0.1$ & $-0.9\pm3.9$ & 0.90 \\
$1<z<2$ & 6 & 132 & 19 & 0 & 157 &  $-1.5\pm0.6$ & Unconstrained & Unconstrained & 1.42 \\
$2<z<3$ & 7 & 82 & 2 & 8 & 99 &  $-1.3\pm1.4$ & $1.9\pm1.9$ & $6.3\pm4.6$ & 1.00 \\
$z>3$   & 7 & 15 & 0 & 2 & 24 &  $-1.4\pm4.0$ &  $-3.7\pm4.3$ & $-2.7\pm10.9$ & 2.35 \\
\hline
All data & 29 & 293 & 21 & 16 & 359 & $-1.6\pm0.5$ & $-0.2\pm0.1$ & $1.7\pm1.3$ & 1.27 \\
\hline
\end{tabular}
\caption{\label{table5}One-dimensional marginalized one-sigma constraints for $\alpha$, $\mu$ an $g_p$, for the various redshift bins as well as for the full dataset for comparison. The first 6 columns show the number of measurements in each of our datasets that fall in each redshift bin. Columns 7-9 show the 1D constraints, given in parts per million. The last column has the reduced chi-square for maximum of the 3D likelihood.}
\end{center}
\end{table*}
%%%%%%%%%%%%%%%%%%%%%%%%%%%%%%%%%%%%%%%%%%%%%%%%%%%%%%%%%%%%%%%%%%%%%%%%%%%%%%%%%%

The results of this tomographic analysis are shown in Figs. \ref{fig4} and \ref{fig5} and in Table \ref{table5}. The values of the reduced chi-square improve, with the only very large one occurring for the $z>3$ bin. Conversely the ones for $0<z<1$ and especially for $2<z<3$ are now quite good. Thus from a purely statistical point of view, the division into redshift bins is certainly warranted. It is worthy of note that the best-fit value for $\alpha$ is almost redshift-independent, although it is better determined at the lower redshifts, $z<2$. As for $\mu$ and $g_p$, they are well constrained for $z<1$ but unconstrained for $1<z<2$: in this redshift bin, current observations can only constrain the combination $g_p/\mu$.

%%%%%%%%%%%%%%%%%%%%%%%%%%%%%%%%%%%%%%%%%%%%%%%%%%%%%%%%%%%%%%%%%%%%%%%%%%%%%%%%%%
\begin{table*}
\begin{center}
\begin{tabular}{|c|cccc|c|c|c|c|}
\hline
Redshift & Tab.I & Tab.II & Tab.III & Total & ${ \Delta\alpha}/{\alpha}$ (ppm) & ${ \Delta\mu}/{\mu}$ (ppm) & ${\Delta g_p}/{g_p}$ (ppm) & $\chi^2_\nu$  \\
\hline
$0<z<1$ & 9 & 0 & 6 & 15 &  $-2.8\pm2.6$ & $-0.2\pm0.1$ & $1.8\pm6.5$ & 1.05 \\
$1<z<2$ & 6 & 19 & 0 & 25 &  $-0.6\pm0.6$ & Unconstrained & Unconstrained & 2.72 \\
$2<z<3$ & 7 & 2 & 8 & 17 &  $-4.3\pm1.7$ & $3.2\pm1.9$ & $14.9\pm5.8$ & 1.09 \\
$z>3$   & 7 & 0 & 2 & 9 &  $-4.2\pm4.4$ &  $-2.6\pm4.3$ & $4.6\pm12.6$ & 4.39 \\
\hline
All data & 29 & 21 & 16 & 66 & $-1.3\pm0.6$ & $-0.2\pm0.1$ & $1.0\pm1.5$ & 2.33 \\
\hline
\end{tabular}
\caption{\label{table6}Same as Table \protect\ref{table5}, but without including the archival measurements of Webb {\it et al}.}
\end{center}
\end{table*}
%%%%%%%%%%%%%%%%%%%%%%%%%%%%%%%%%%%%%%%%%%%%%%%%%%%%%%%%%%%%%%%%%%%%%%%%%%%%%%%%%%

A pertinent question is whether the above results are driven by the archival measurements of Webb {\it et al.}, which despite the growing number of recent measurements still comprise more than eighty percent of our full dataset. To address this question, Table \ref{table6} repeats the analysis of Table \ref{table5} including only the 66 dedicated measurements. Again we note the reasonable values of the reduced chi-square for the redshift bins $0<z<1$ and $2<z<3$, while those of the other two bins (and also the one for the full set of 66 measurements) are very high. This therefore suggests that some of the measurements in the Table I dataset have unreliable (that is, too small) uncertainties, with the problem presumably being worse at the highest redshifts.

Given the caveats of the previous paragraph no strong conclusions can be drawn from this analysis. Nevertheless, if one takes the data at face value, it is interesting that the  Webb {\it et al.} data does enhance the preference for a negative $\alpha$ in the redshift range $1<z<2$. On the other hand, and somewhat counterintuitively, it leads to a statistical preference for a less negative $\alpha$ for $z>2$. This is the result of the fact that the direct $\mu$ measurements in the $2<z<3$ bin slightly prefer a positive $\mu$; then without the  Webb {\it et al.} data, a negative $\alpha$ and a positive $g_p$ are preferred by the Table I data. In any case, our results highlight the point that combined measurements with improved sensitivities play an important role in testing the consistency of direct $\alpha$ and $\mu$ measurements.

%%%%%%%%%%%%%%%%%%%%%%%%%%%%%%%%%%%%%%%%%%%%%%%%%%%%%%%%%%%%%%%%%%%%%%%%%%%%%%
\section{\label{pqrs}Relating the variations of different couplings}

So far we have treated the possible relative variations of the different couplings as independent parameters. However, we should note that in most (if not all) well-motivated models all such variations will be related---although the specific relations will be highly model-dependent. For example, in a generic class of unification scenarios discussed in \cite{Campbell,Coc,Luo} the relative variations of these parameters are related via
\begin{equation}
\frac{\Delta \mu}{\mu} = [0.8 R - 0.3 (1+S)] \frac{\Delta \alpha}{\alpha}
\end{equation}
\begin{equation}
\frac{\Delta g_p}{g_p} = [0.1 R - 0.04(1+S)] \frac{\Delta \alpha}{\alpha}
\end{equation}
where the parameter $R$ is related to the strong sector of the model while $S$ is related to the electroweak one. Constraints on these parameters have been obtained in previous works  \cite{Frigola,Ferreira2015,Clocks}, though one finds that current data only constraints a particular combination of $R$ and $S$ .

%%%%%%%%%%%%%%%%%%%%%%%%%%%%%%%%%%%%%%%%%%%%%%%%%%%%%%%%%%%%%%%%%%%%%%%%%%%%%%%%%%
\begin{figure}
\begin{center}
\includegraphics[width=2.6in]{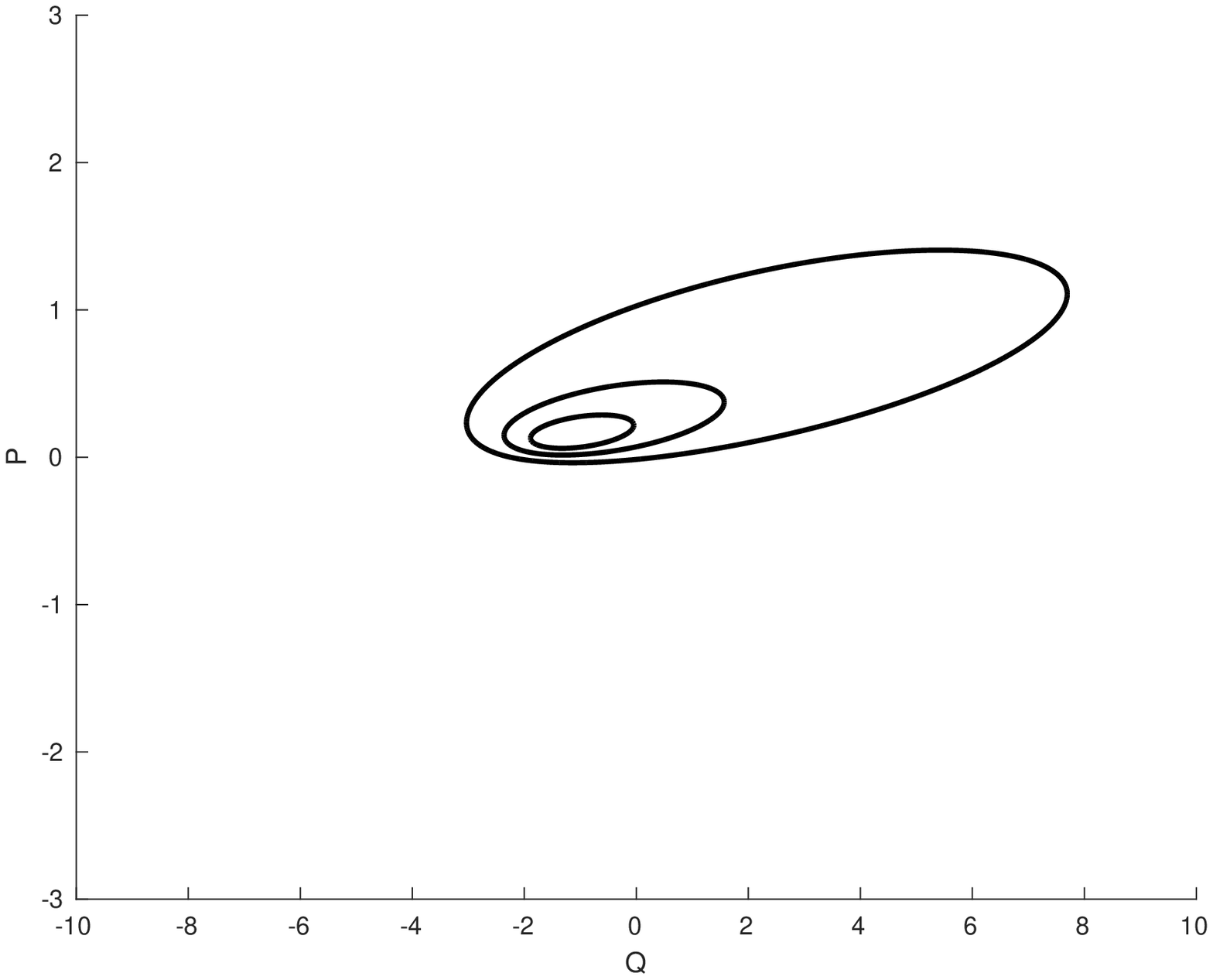}
\vskip0.25in
\includegraphics[width=2.6in]{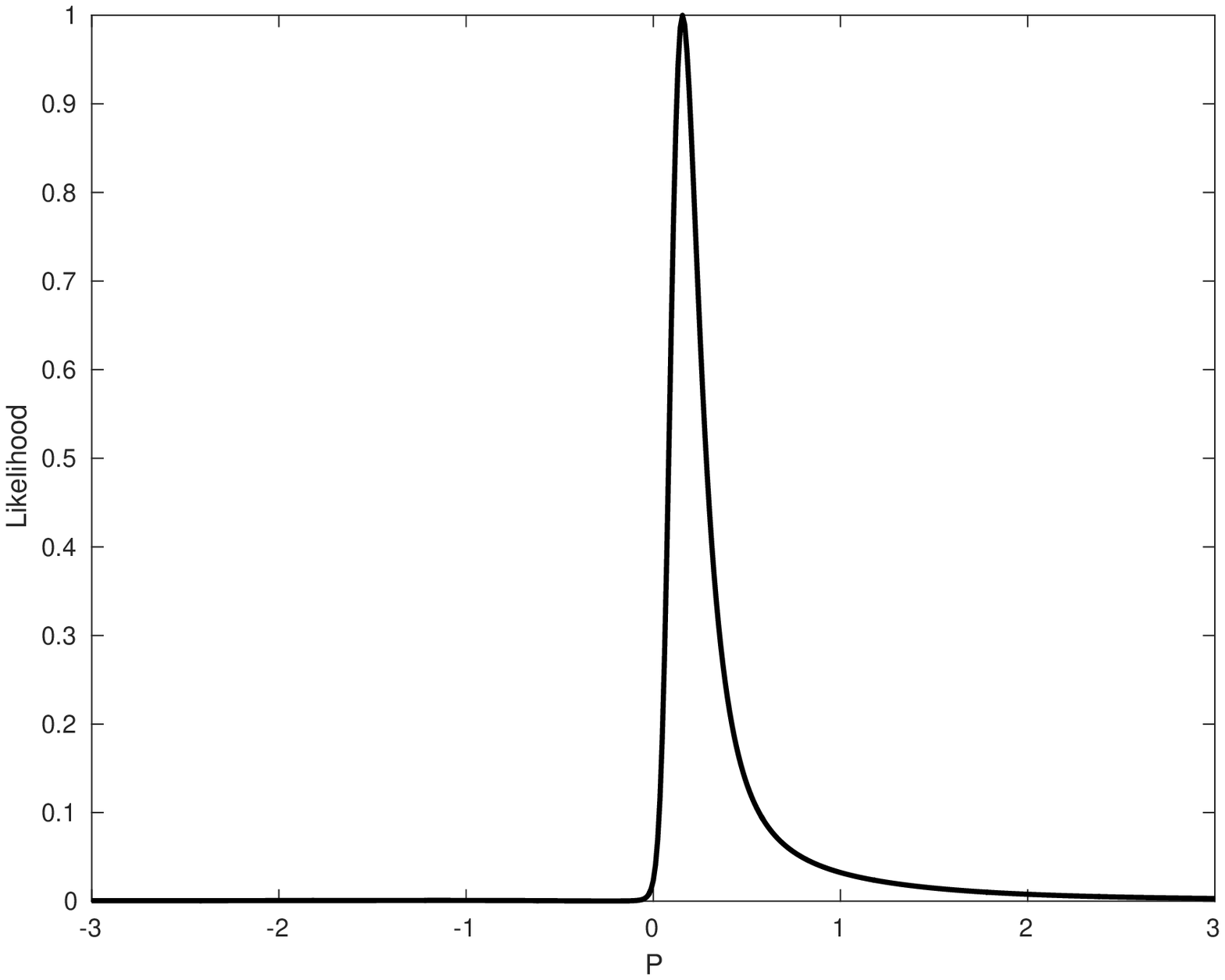}
\vskip0.25in
\includegraphics[width=2.6in]{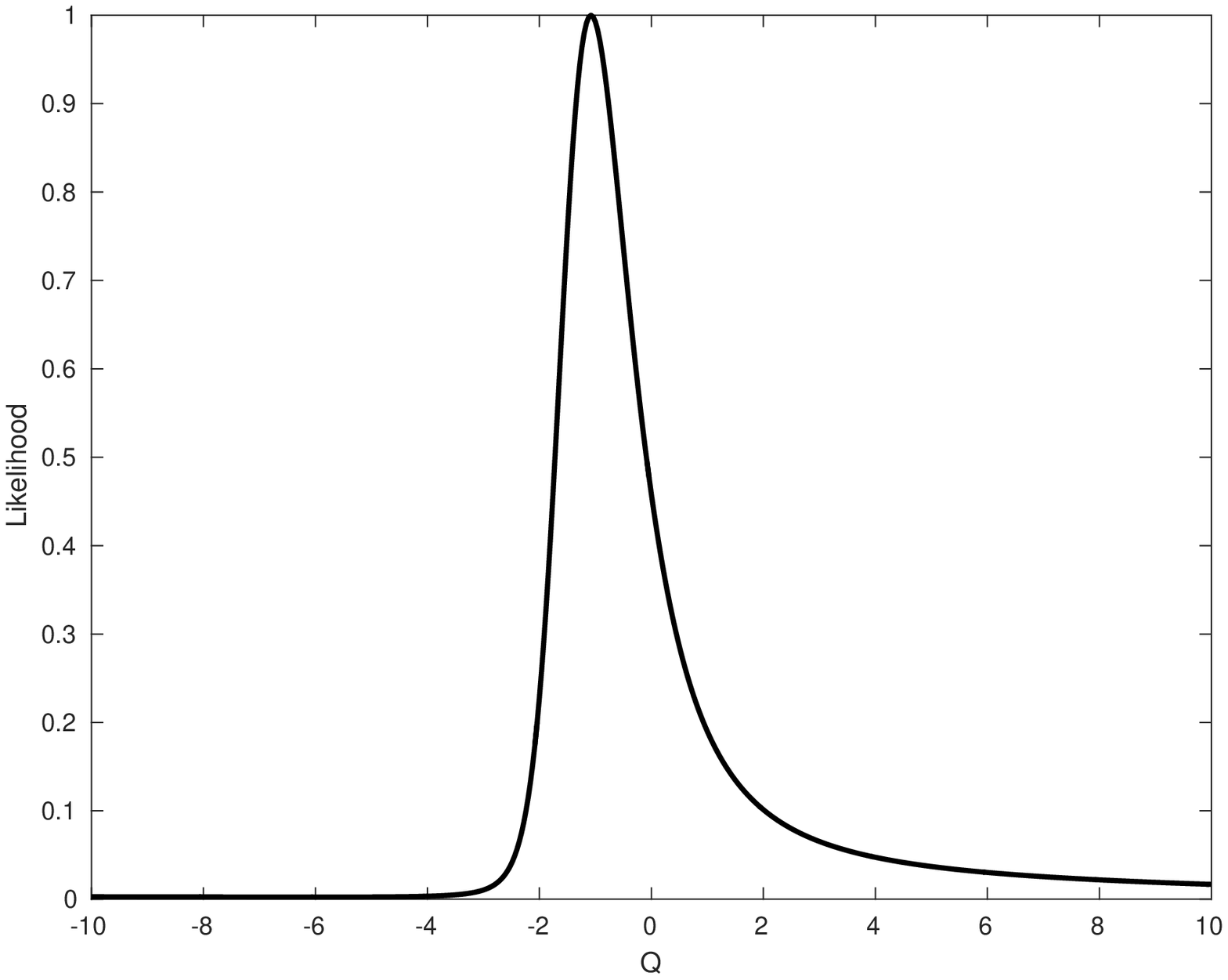}
\end{center}
\caption{\label{fig6}Two-dimensional likelihood in the P--Q parameter space (top panel; one, two and three sigma contours), and 1D likelihood for each parameter with the other marginalized (middle and bottom panels), for our full dataset.}
\end{figure}
%%%%%%%%%%%%%%%%%%%%%%%%%%%%%%%%%%%%%%%%%%%%%%%%%%%%%%%%%%%%%%%%%%%%%%%%%%%%%%%%%%

Here we will take a simpler and perhaps more intuitive phenomenological approach, defining
\begin{equation}
\frac{\Delta \mu}{\mu} = P \frac{\Delta \alpha}{\alpha}
\end{equation}
\begin{equation}
\frac{\Delta g_p}{g_p} = Q\frac{\Delta \alpha}{\alpha}
\end{equation}
and obtaining constraints on the $(P,Q)$ plane. Naturally these can be related to those on the $(R,S)$ plane, since
\begin{equation}
R=10(2P-15Q)
\end{equation}
\begin{equation}
(1+S)=50(P-8Q)\,.
\end{equation}
Note that in the unification models under consideration R and S are universal (redshift-independent) parameters, and we therefore assume that this is also the case for our more phenomenological parameters P and Q. Hence we can jointly use all our datasets to constrain them.

These constraints are shown in Fig. \ref{fig6}. We note that there are several examples of specific models studied in the literature for which P is (in absolute value) one or two orders of magnitude larger than unity. For example Coc {\it et al.} \cite{Coc} suggest typical values of $R\sim36$ and $S\sim160$, leading to $P\sim-20$, while in the dilaton-type model studied by Nakashima {\it et al.} \cite{Nakashima} we have $R\sim109$ and $S\sim0$, and thus $P\sim87$. Additional discussion can be found in the review by Uzan \cite{Uzan}. The extent to which this is a generic property of all unification models is at present unclear. The current data leads to
\begin{equation}
P=0.16^{+0.10}_{-0.07}
\end{equation}
\begin{equation}
Q=-1.1^{+0.8}_{-0.6}\,,
\end{equation}
both at the one sigma ($68.3\%$) confidence level. For comparison, a joint analysis of atomic clock measurements in \cite{Clocks} leads to a a constraint that in terms of the parameter P reads $P=1.5\pm4.5$.

The tight low-redshift measurements of $\mu$ constrains P to be not much larger than unity (even at the three sigma level). Given that the data slightly prefers a negative value of $\alpha$, the combination with the measurements of Tables III and I leads to preferred values for P and Q that are respectively positive and negative. Given the aforementioned caveats on these datasets (especially concerning Table I), these constraints should be interpreted with some caution, but again they highlight the potential constraining power of these measurements.

%%%%%%%%%%%%%%%%%%%%%%%%%%%%%%%%%%%%%%%%%%%%%%%%%%%%%%%%%%%%%%%%%%%%%%%%%%%%%%
\section{\label{dipvar}Spatial variations}

The analysis by Webb {\it et al.}  of their large archival dataset provided evidence for spatial variations of the fine-structure constant, $\alpha$, at the level of a few parts per million (ppm) \cite{Dipole,KingDip}. Both their analysis and those of subsequent works \cite{tests1,tests2,tests3,tests4} find evidence for a spatial dipole at a statistical level of significance of more than four standard deviations. These previous studies were restricted to the archival dataset. A recent analysis \cite{Pinho0}, combining this with the then-existing set of 11 dedicated measurements found that the dipole was still a good fit, although the preferred amplitude was reduced by twenty percent. We now update this analysis, given that there are now 21 dedicated $\alpha$ measurements in Table II, significantly increasing the sky coverage, and that some of the previously existing measurements have been improved.

We note that the third measurement listed in Table II is the weighted average from measurements along three lines of sight that are widely separated on the sky, HE1104-1805A, HS1700+6416 and HS1946+7658, reported in \cite{Songaila}. (The authors only report this average and not the individual measurements, and our attempts to directly contact these authors were unsuccessful.) For this reason we have listed the result in Table II for completeness (and used it for the analyses reported in the previous sections) but naturally it will not be included in our spatial variations analysis. For this purpose the more recent dataset therefore has 20 different measurements.

We will fit the $\alpha$ measurements to two different phenomenological parametrizations. The first is a pure spatial dipole for the relative variation of $\alpha$
\begin{equation}\label{puredipole}
\frac{\Delta\alpha}{\alpha}(A,\Psi)=A\cos{\Psi}\,,
\end{equation}
which depends on the orthodromic distance $\Psi$ to the North Pole of the dipole (the locus of maximal positive variation) given by
\begin{equation}\label{ortho}
\cos{\Psi}=\sin{\theta_i}\sin{\theta_0}+\cos{\theta_i}\cos{\theta_0}\cos{(\phi_i-\phi_0)}\,,
\end{equation}
where $(\theta_i,\phi_i)$ are the Declination and Right Ascension of each measurement and $(\theta_0,\phi_0)$ those of the North Pole. These latter two coordinates, together with the overall amplitude $A$, are our free parameters. This parametrizations has been considered in all the aforementioned previous studies and thus provides a validation test of our own analysis. We note that we do not consider an additional monopole term, both because there is no strong statistical preference for it in previous analyses \cite{Dipole,KingDip} and because it is physically clear that any such term would be understood as being due to the assumption of terrestrial isotopic abundances, in particular of Magnesium---we refer the interested reader to \cite{Monopole} for a detailed technical discussion of this point.

%%%%%%%%%%%%%%%%%%%%%%%%%%%%%%%%%%%%%%%%%%%%%%%%%%%%%%%%%%%%%%%%%%%%%%%%%%%%%%%%%%
\begin{figure}
\begin{center}
\includegraphics[width=2.6in]{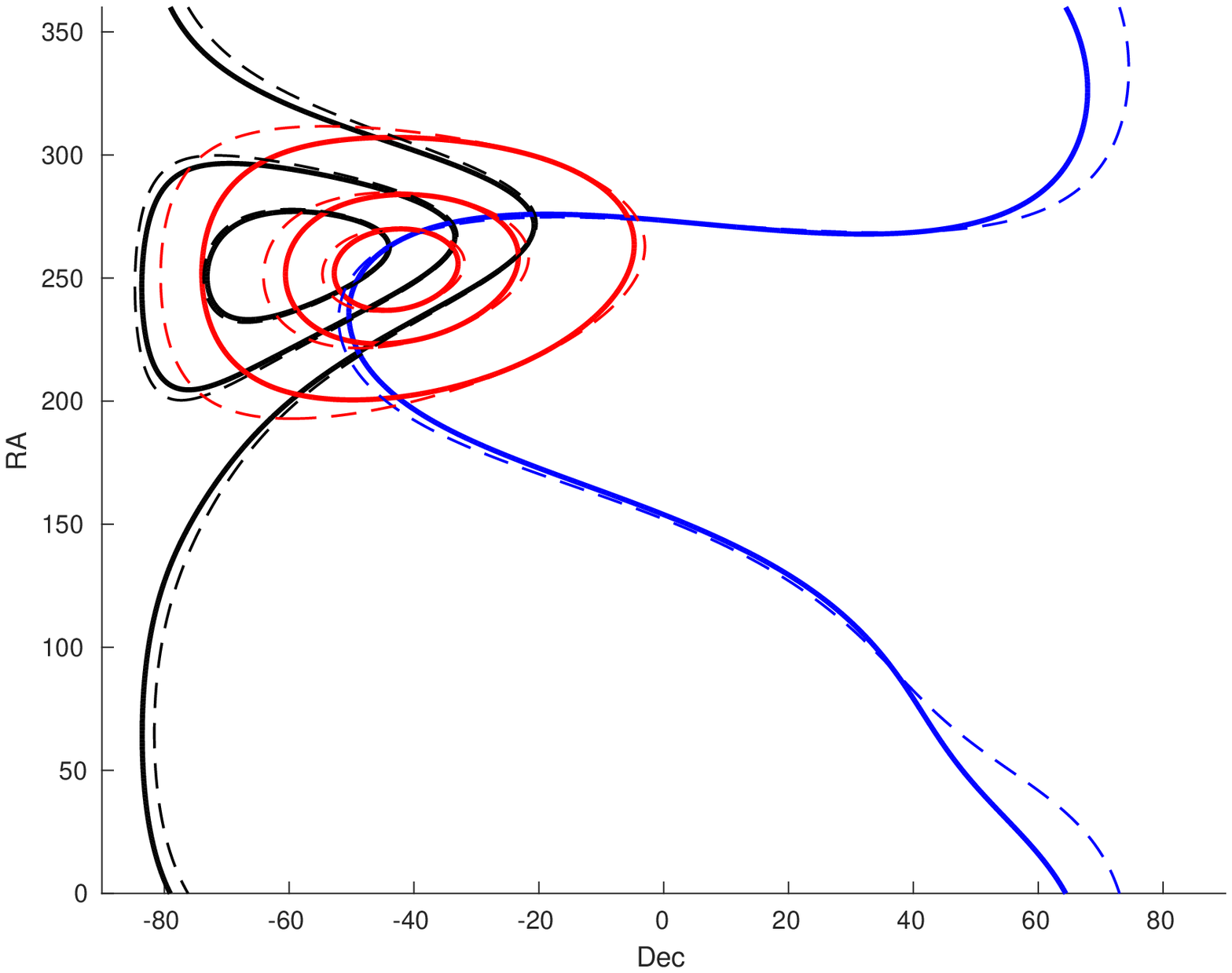}
\vskip0.25in
\includegraphics[width=2.6in]{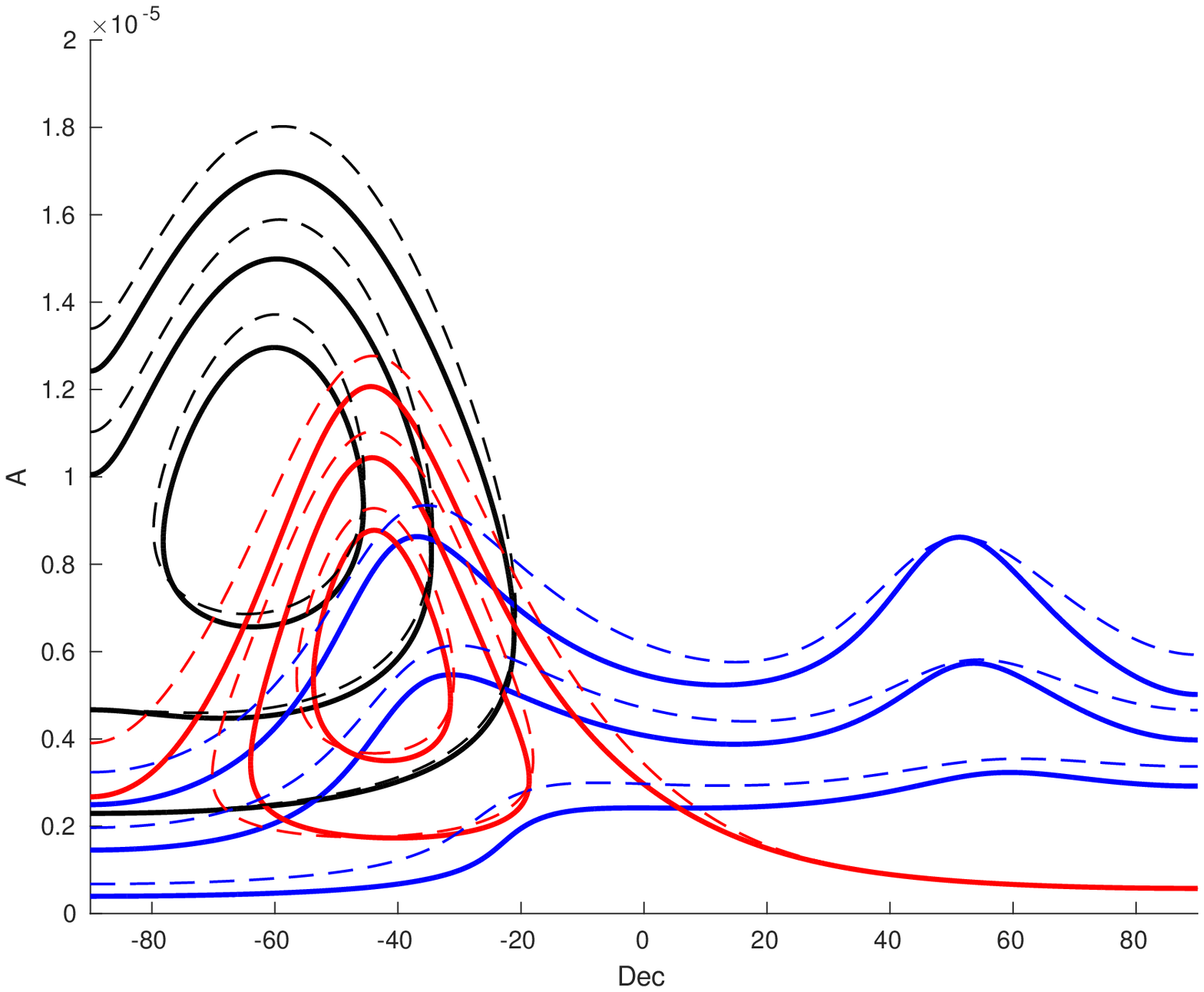}
\vskip0.25in
\includegraphics[width=2.6in]{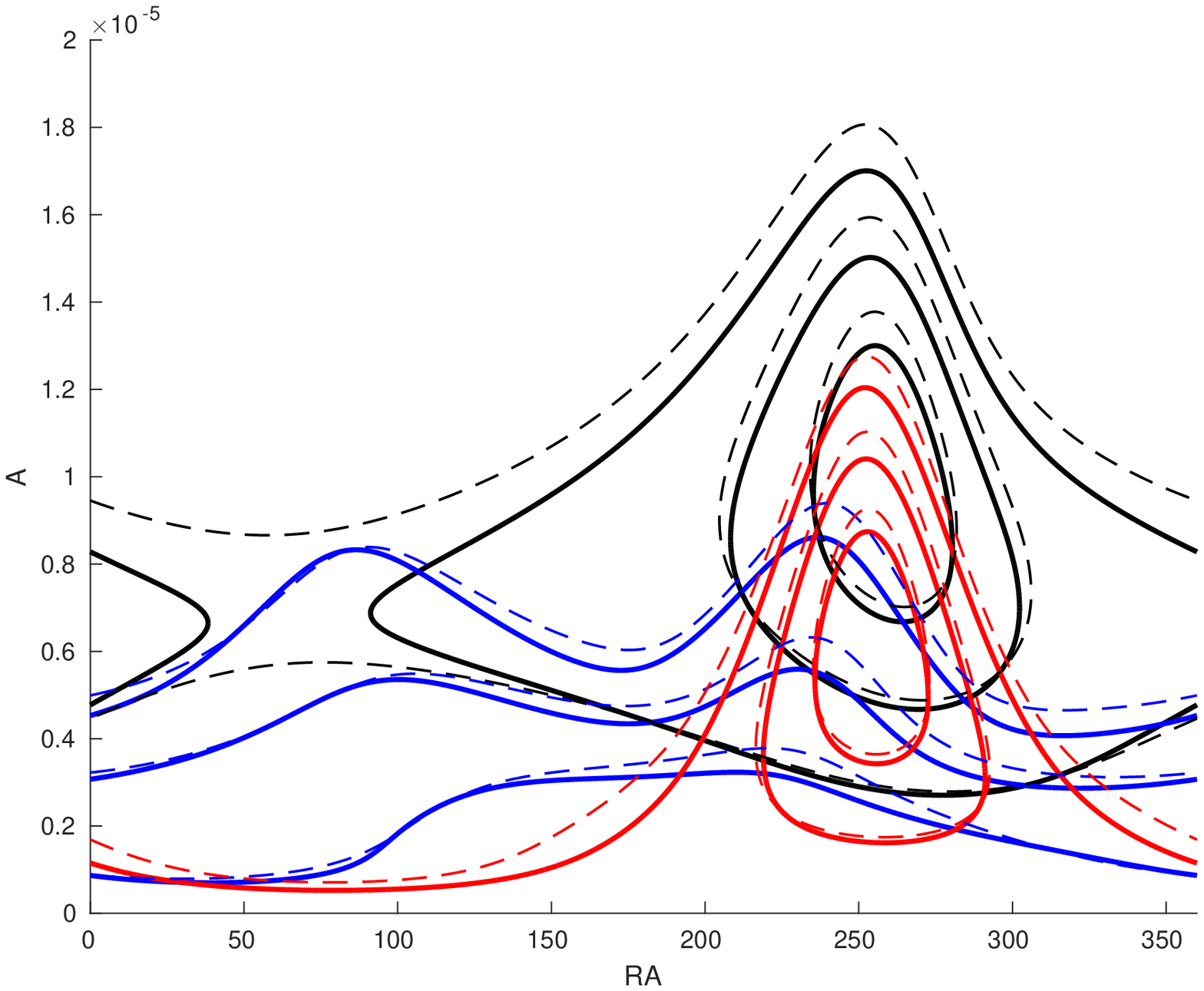}
\end{center}
\caption{\label{fig7}Two-dimensional likelihood, with the third parameter marginalized, for a pure spatial dipole (solid) and for a dilaton-like redshift dependence (dashed), for the Webb {\it et al.} data (black lines), the Table II data (blue lines) and the combined data (red lines). One, two and three sigma contours are shown in all cases.}
\end{figure}
%%%%%%%%%%%%%%%%%%%%%%%%%%%%%%%%%%%%%%%%%%%%%%%%%%%%%%%%%%%%%%%%%%%%%%%%%%%%%%%%%%
\begin{figure}
\begin{center}
\includegraphics[width=2.6in]{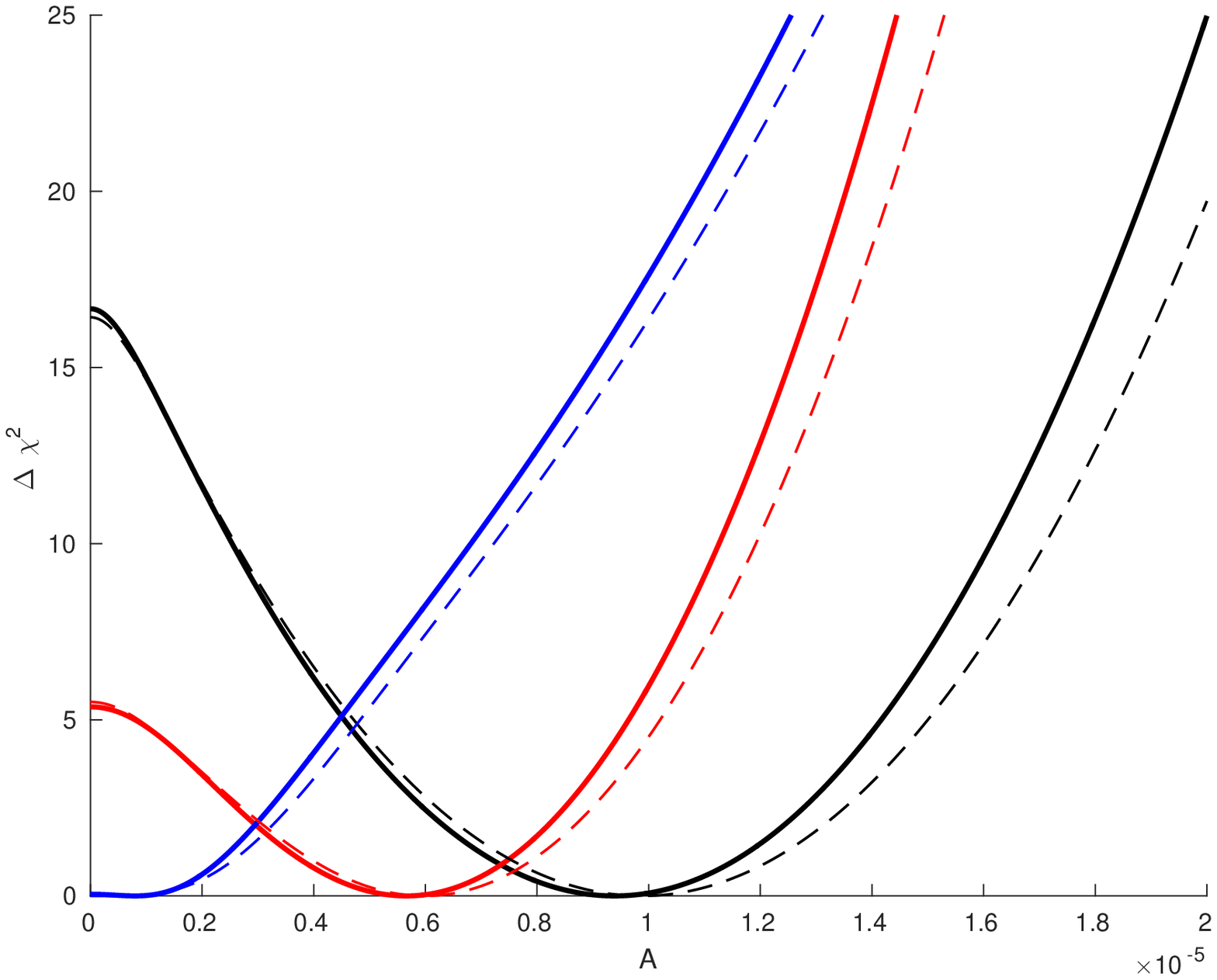}
\vskip0.25in
\includegraphics[width=2.6in]{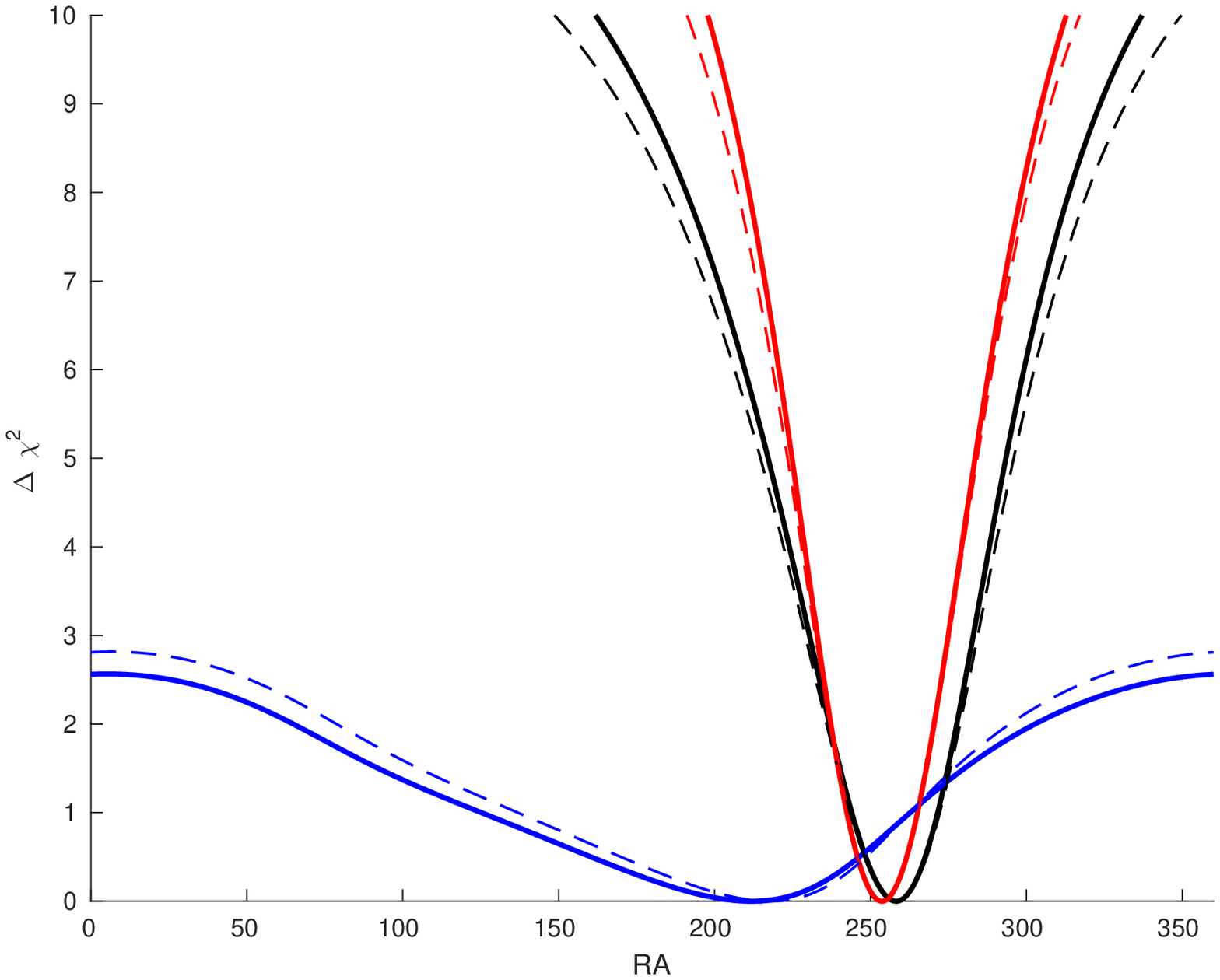}
\vskip0.25in
\includegraphics[width=2.6in]{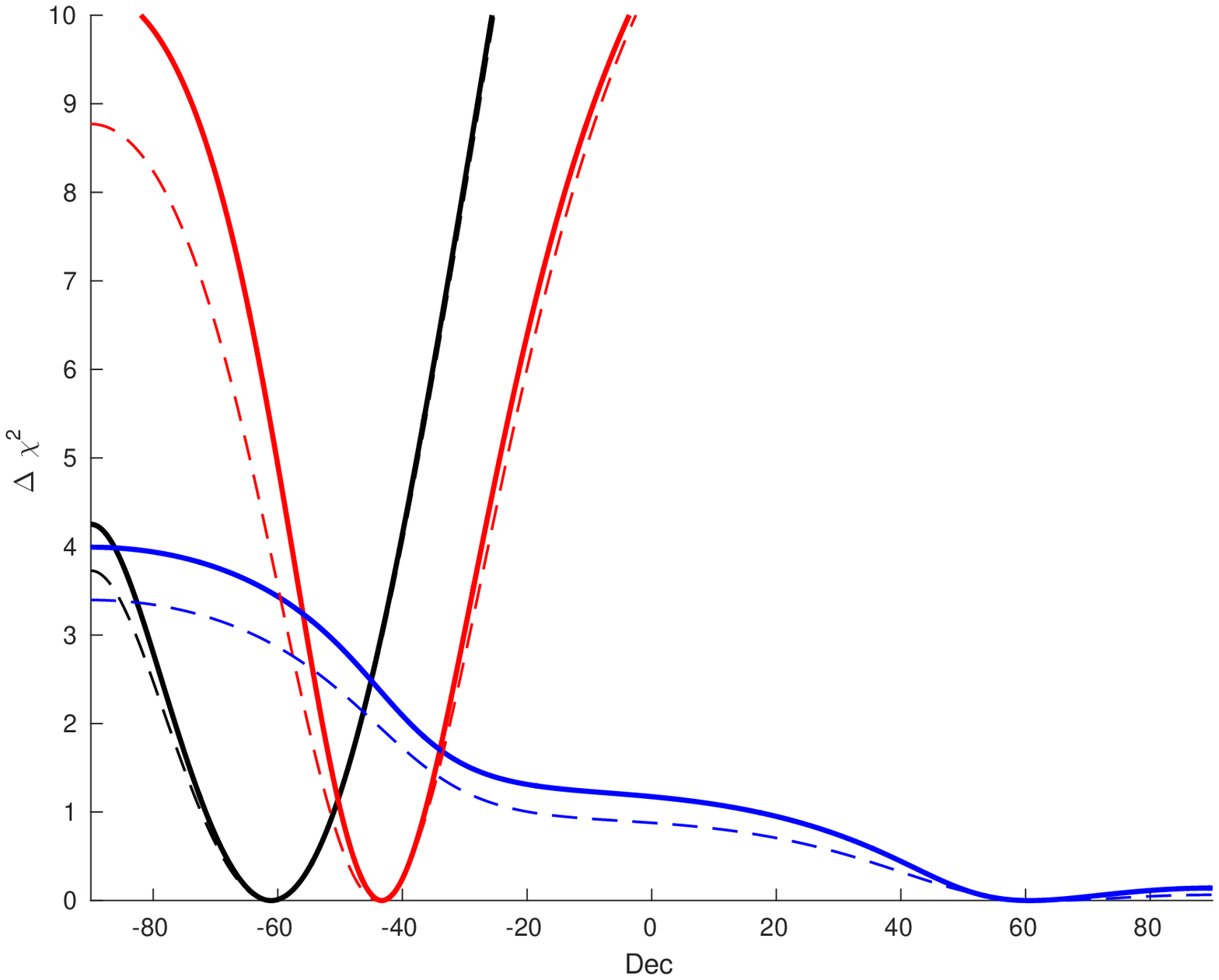}
\end{center}
\caption{\label{fig8}1D likelihood (with the other two parameters marginalized) for the same models and datasets of Fig. \protect\ref{fig7}.}
\end{figure}
%%%%%%%%%%%%%%%%%%%%%%%%%%%%%%%%%%%%%%%%%%%%%%%%%%%%%%%%%%%%%%%%%%%%%%%%%%%%%%%%%%

We will also consider a parametrization where there is an implicit time dependence in addition to the spatial variation. Previous analyses considered the case of a dependence on look-back time, but this requires the assumption of a cosmological model and moreover it is not clear how such a dependence would emerge from realistic varying $\alpha$ models. We will instead assume a logarithmic dependence on redshift
\begin{equation}\label{redshiftdipole}
\frac{\Delta\alpha}{\alpha}(A,z,\Psi)=A\, \ln{(1+z)}\, \cos{\Psi}\,.
\end{equation}
This has the practical advantage of not requiring any additional free parameters, but such dependencies are also typical of dilaton-type models \cite{Dilaton}. As in previous analyses, this parametrization is mainly considered as a means to assess the ability of the data to discriminate between different models for spatial variations.

%%%%%%%%%%%%%%%%%%%%%%%%%%%%%%%%%%%%%%%%%%%%%%%%%%%%%%%%%%%%%%%%%%%%%%%%%%%%%%%%%%
\begin{table*}
\begin{center}
\begin{tabular}{|c|c|c|c|}
\hline
Dataset \& C.L. & Amplitude ($ppm$) & Right Ascension ($h$) & Declination (${}^\circ$) \\
\hline
Webb {\it et al.} ($68.3\%$) & $9.4\pm2.2$ & $17.2\pm1.0$ & $-61\pm10$ \\
Webb {\it et al.} ($99.7\%$) & $9.4\pm6.4$ & $17.2^{+4.4}_{-5.3}$ & $<-28$ \\
\hline
Table II ($68.3\%$) & $<2.3$ & $14.1^{+3.4}_{-5.8}$ & $>17$ \\
Table II ($99.7\%$) & $<6.4$ & N/A & N/A \\
\hline
All data ($68.3\%$) & $5.6\pm1.8$ & $16.9\pm0.8$ & $-43\pm7$ \\
All data ($99.7\%$) & $<10.9$ & $16.9^{+3.4}_{-3.2}$ & $-43^{+34}_{-31}$ \\
\hline
\end{tabular}
\caption{\label{table7}One- and three-sigma constraints on the Amplitude and coordinates of maximal variation (Right Ascension and Declination) for a pure spatial dipole variation of $\alpha$. The 'All Data' case corresponds to using the data of Webb {\it et al.} \cite{Dipole} together with the 20 individual measurements presented in Table \protect\ref{table2}.}
\end{center}
\end{table*}
%%%%%%%%%%%%%%%%%%%%%%%%%%%%%%%%%%%%%%%%%%%%%%%%%%%%%%%%%%%%%%%%%%%%%%%%%%%%%%%%%%

Our results are shown in Figs. \ref{fig7} and \ref{fig8} and in Table \ref{table7}. For the Webb {\it et al.} data we recover the statistical preference for a dipole at more than four standard deviations, while there is no preference for a dipole in the more recent data. Combining the two datasets, the statistical preference for a dipole is reduced to only 2.3 standard deviations, and the best-fit amplitude is less than 6 ppm. As for the direction of maximal variation, we note that the preferred Declination is significantly changed, moving by about 18 degrees, while the Right Ascension is comparatively less affected. This information is useful for the purpose of selecting targets for future observations.

Comparing our results for the pure spatial dipole and the redshift-dependent one, we see that they are very similar (with the constraints on the latter being very slightly weaker). This is visually clear in Figs. \ref{fig7} and \ref{fig8} (where the results for both models are represented), and for this reason Table \ref{table7} only reports the results for the pure spatial case. The current sensitivity and redshift distribution of the measurements is not sufficient to distinguish between these models.

\begin{figure}[t!]
\begin{center}
\includegraphics[width=2.6in]{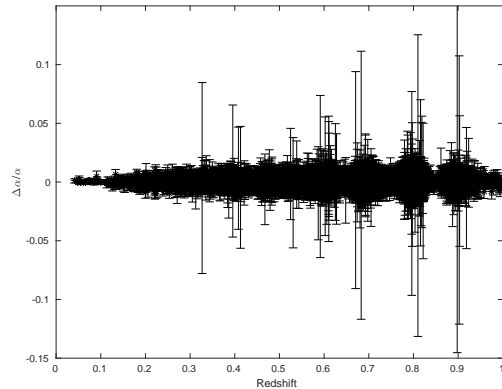}
\end{center}
\caption{\label{fig9}The sample of 13175 emission line measurements of $\alpha$ from the SDSS-III/BOSS DR12 quasar sample of Albareti {\it et al.} \protect\cite{Albareti}. The weighted mean of the measurements is $9\pm18$ ppm.}
\end{figure}

An additional independent test of possible spatial variations can be done with the sample of 13175 emission line measurements of $\alpha$ from the SDSS-III/BOSS DR12 quasar sample of Albareti {\it et al.} \protect\cite{Albareti}. While the sensitivity of each of their individual measurements of the relative variation of $\alpha$ is much worse than the ones in Table II (ranging from $2.4\times10^{-4}$ to $1.5\times10^{-2}$, cf. Fig. \ref{fig9}, to be compared to parts-per-million), the large number of measurements covering a significant fraction of the sky still allows for a worthwhile test of spatial variations. For comparison, the weighted mean of the 13175 measurements, which span the redshift range $0.041<z<0.997$ is
\begin{equation}\label{albaretimean}
\frac{\Delta\alpha}{\alpha}=9\pm18 \,ppm\,;
\end{equation}
the original work \protect\cite{Albareti} also provides weighted means for sub-samples in various redshift bins, but since our purpose is to test for possible spatial variations we will use the individual measurements and their directions on the sky. For the reasons already explained we will restrict our analysis to the case of a pure spatial dipole.

%%%%%%%%%%%%%%%%%%%%%%%%%%%%%%%%%%%%%%%%%%%%%%%%%%%%%%%%%%%%%%%%%%%%%%%%%%%%%%%%%%

%%%%%%%%%%%%%%%%%%%%%%%%%%%%%%%%%%%%%%%%%%%%%%%%%%%%%%%%%%%%%%%%%%%%%%%%%%%%%%%%%%
\begin{figure}
\begin{center}
\includegraphics[width=2.6in]{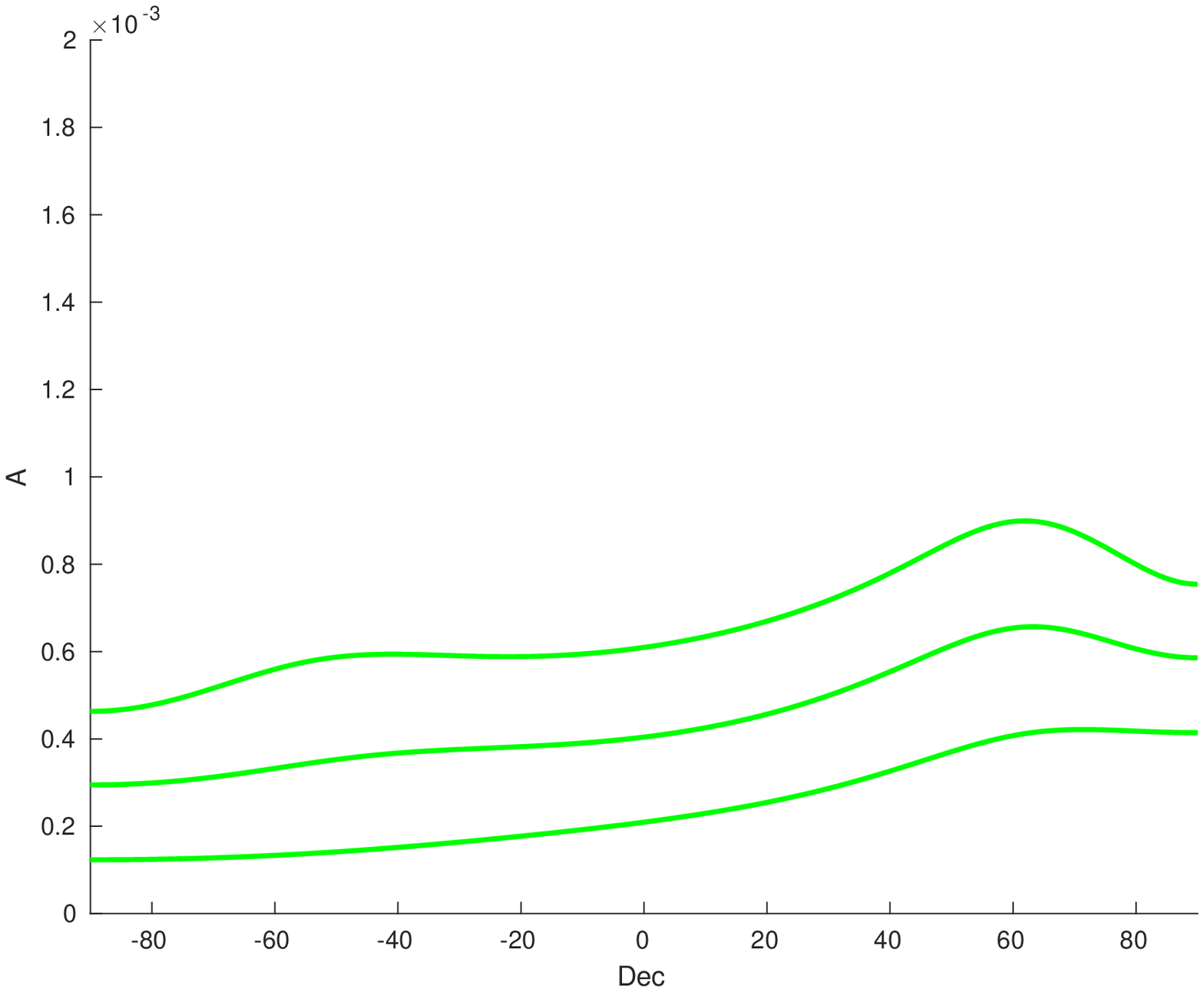}
\vskip0.25in
\includegraphics[width=2.6in]{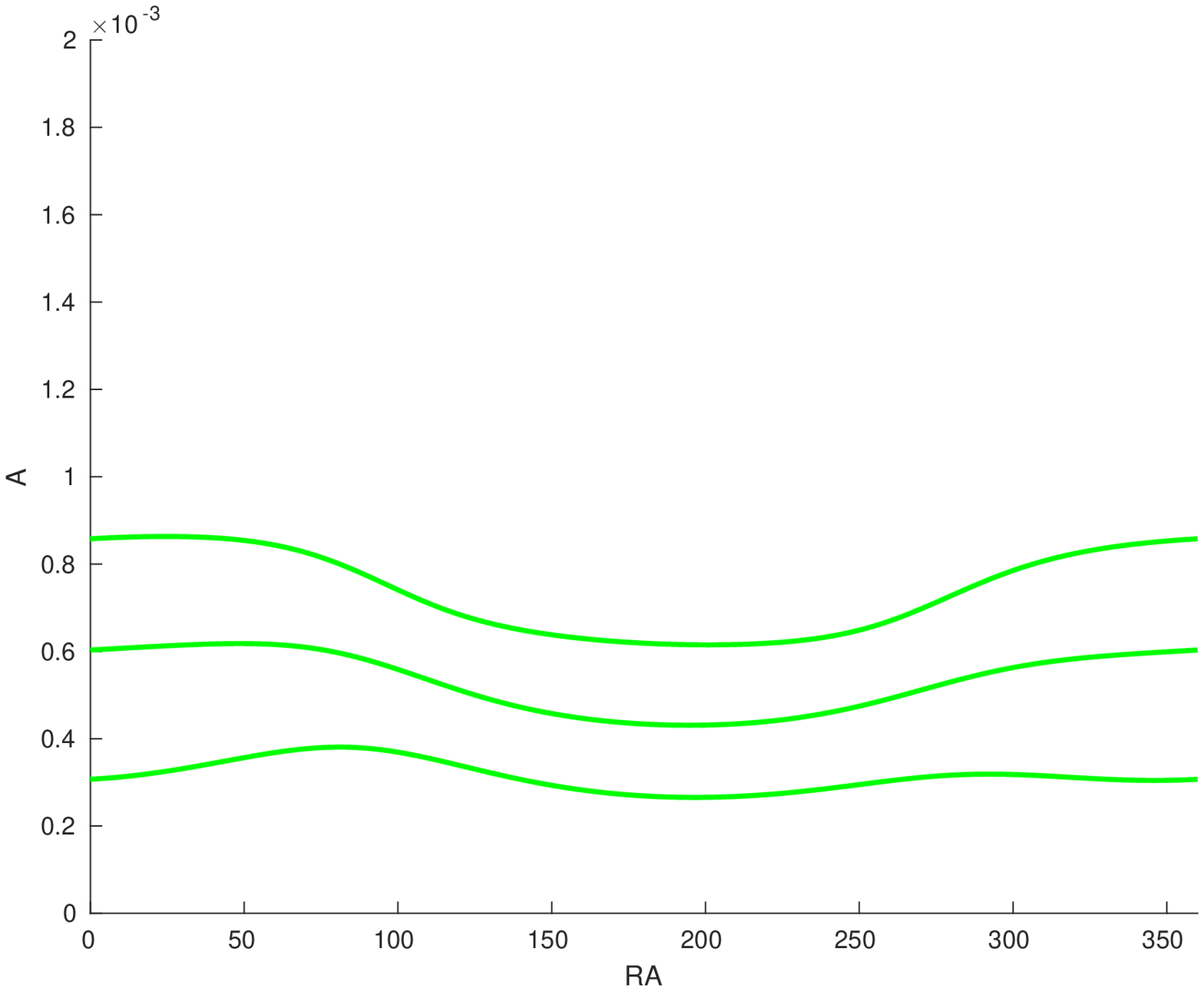}
\vskip0.25in
\includegraphics[width=2.6in]{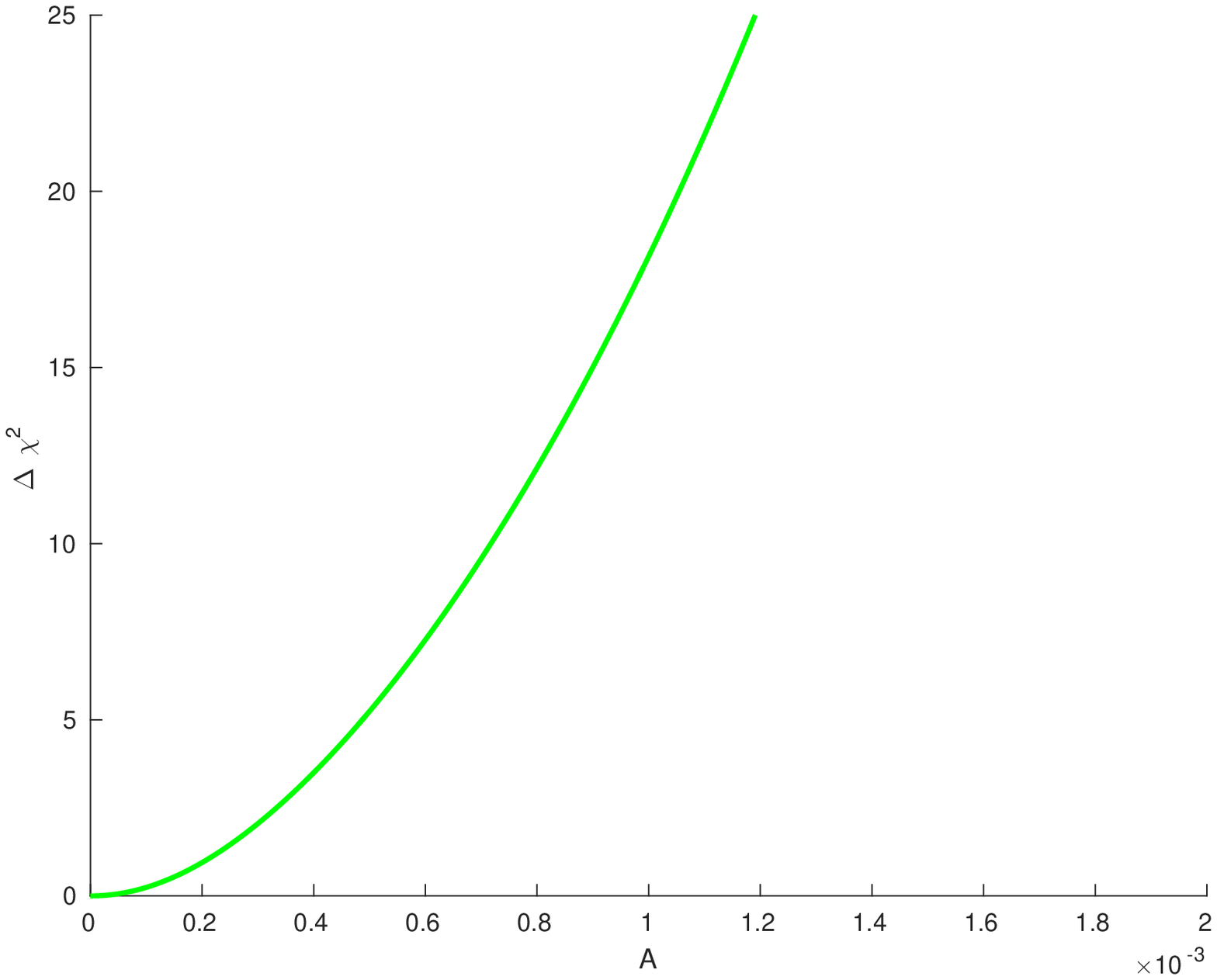}
\end{center}
\caption{\label{fig10}Relevant 2D (one, two and three sigma confidence levels) and 1D likelihood constraints on a pure spatial dipole the Albareti {\it et al.} data.}
\end{figure}
%%%%%%%%%%%%%%%%%%%%%%%%%%%%%%%%%%%%%%%%%%%%%%%%%%%%%%%%%%%%%%%%%%%%%%%%%%%%%%%%%%

The results are summarized in Fig. \ref{fig10}: there is no preference for a particular direction on the sky, and we obtain the following three-sigma upper bound for the amplitude of a putative dipole
\begin{equation}\label{sdss}
A_{SDSS}<7\times 10^{-4} \, (99.7\% C.L.)\,.
\end{equation}
This bound is about 64 times weaker than the one obtained above from the absorption line measurements, but it is independent from it. Moreover, it is stronger than recent bounds on spatial variations coming from the combination of Sunyaev-Zel'dovich cluster measurements and Planck satellite data (and even stronger than analogous bounds from the Planck cosmic microwave background alone) \cite{SZclusters}.

%%%%%%%%%%%%%%%%%%%%%%%%%%%%%%%%%%%%%%%%%%%%%%%%%%%%%%%%%
\section{\label{concl}Conclusions}

In this work we have fully updated earlier analyses \cite{Frigola,Ferreira2015} of the consistency of currently available astrophysical tests of the stability of fundamental couplings. The rapid development of the field, with new and improved measurements frequently appearing---especially those of the fine-structure constant $\alpha$---warrant an updated analysis, and as our results show the new measurements do have a significant impact. Our full dataset comprised 359 measurements of $\alpha$, $\mu$ as well as several of their combinations (also including $g_p$). These span the redshift range  $0.2<z<6.4$ and also a broad range of sensitivities, from about 0.1 ppm to more than 100 ppm. We also considered the large datasets of absorption line measurements by Webb {\it et al.} \cite{Dipole} and, when constraining spatial variations, the set of recent emission line measurements by Albareti {\it et al.} \cite{Albareti}.

Overall, our analysis suggests that there are currently no robust indications of time or space variations. Some preferences for variations at about the two-sigma level of statistical significance do exist, but it is presently unclear what their origin is. Specifically, the results tend to be different at low and high redshifts. While this could indicate different behaviors in the matter era and the recent acceleration era, it could also point to hidden systematics, since radio/mm measurements are typically done at lower (median) redshifts than the optical/UV ones.

Clearly, the main open question concerning the current data is the extent to which systematic errors have been accounted for. In the case of optical/UV measurements, possible sources for these have recently been studied in some detail \cite{LP1,LP2,LP3,Syst}, but the extent to which one can model them and correct them {\it a posteriori} is still a subject of ongoing study and discussion in the community. Our analysis also suggests that uncertainties in the combined measurements of Table I may be underestimated.

Clarifying these issues is essential. One way forward is to find lines of sight where these measurements can be carried out both in the optical and in the radio bands, but the number of such ideal targets is likely to be small. An easier goal is to extend the range of radio/mm measurements so that they overlap with the ones in the optical/UV---this would provide an important way to characterize hidden systematics. Efforts along these lines are under way, using APEX and ALMA.

Meanwhile, the imminent arrival of the ESPRESSO spectrograph \cite{Espresso}, due for commissioning at the combined Coud\'e focus of the VLT in 2017, will significantly improve the statistical uncertainty and the control over systematics in optical measurements, especially of the fine-structure constant $\alpha$. In addition to the intrinsic importance of these more precise measurements, which our present work highlights, they will also lead to improved constraints on dynamical dark energy and on Weak Equivalence Principle violations. A roadmap for these tests can be found in \cite{grg}.

\begin{acknowledgments}

We thank Jeremy Darling for providing the individual measurements of ${\alpha^{2}g_{p}/\mu}$ of reference \cite{DarlingNew}, and Franco Albareti and Ana Catarina Leite for helpful discussions on the subject of this work.

This work was done in the context of project PTDC/FIS/111725/2009 (FCT, Portugal), with additional support from grant UID/FIS/04434/2013. CJM is supported by an FCT Research Professorship, contract reference IF/00064/2012, funded by FCT/MCTES (Portugal) and POPH/FSE (EC).

\end{acknowledgments}

\bibliography{qso}

\end{document}